\documentclass{article}
\usepackage{geometry}
\usepackage[utf8]{inputenc}
\usepackage{cite}
\usepackage{url}
\usepackage{amssymb,amsmath,amsthm}
\usepackage{bm}
\usepackage{graphicx}
\usepackage{enumerate}
\usepackage{microtype}
\usepackage{color}
\usepackage{hyperref}
\usepackage{multirow}
\usepackage{array}
\usepackage{subcaption}
\usepackage{booktabs}

\newcolumntype{C}[1]{>{\centering\let\newline\\\arraybackslash\hspace{0pt}}m{#1}}

\usepackage[ruled,linesnumbered,vlined]{algorithm2e}

\newtheorem{remark}{Remark}

\title{Adaptive Volumetric Parameterization of Simply Connected 3-Manifolds with Applications} 
\author{Zhiyuan Lyu$^{1}$, Qiguang Chen$^{1}$, Lok Ming Lui$^{1}$, Gary P. T. Choi$^{1,\ast}$\\
\\
\footnotesize{$^{1}$Department of Mathematics, The Chinese University of Hong Kong}\\
\footnotesize{$^\ast$To whom correspondence should be addressed; E-mail: ptchoi@cuhk.edu.hk}
}
\date{}

\begin{document}

\maketitle

\begin{abstract}
Volumetric parameterization, the process of mapping a 3-manifold onto a simplified volumetric domain, is important for many tasks in computer graphics and imaging science. However, most prior volumetric parameterization approaches have only utilized standardized domains such as a solid ball regardless of the overall shape of the given 3-manifolds, which introduces significant geometric distortion and affects the subsequent shape processing and analysis tasks. To overcome this issue, in this work we propose a novel volumetric parameterization framework for simply connected 3-manifolds. Specifically, the proposed framework jointly controls local shape and mass distortions, while adapting the target domain during the optimization process. It enables three progressively more flexible target-domain settings for the parameterization: a prescribed solid ellipsoid, a volume-normalized adaptive ellipsoid with variable radii, and a sea-embedded free-boundary domain. For each setting, the parameterization algorithm consists of a 3D quasi-conformality shape update, a diffusion-based density-equalizing update, and a geometric correction procedure for removing element foldings, thereby allowing for volumetric parameterizations with different desired effects. Experimental results are presented to demonstrate the effectiveness of our proposed framework. Moreover, our framework can be easily applied to multiresolution and localized adaptive volumetric remeshing, volumetric registration, and volumetric morphing. Altogether, our work provides a new way for the representation, processing, and analysis of 3-manifolds.

\end{abstract}

%%%%%%%%%%%%%%%%%%%%%%%%
\section{Introduction} \label{sect:introduction}
Shape parameterization, the process of mapping a complicated shape onto a simpler domain, has been widely studied in the areas of geometry processing, graphics, and imaging science~\cite{sheffer2007mesh, hormann2008mesh}. Over the past few decades, a large variety of works have been devoted to surface parameterization. In particular, conformal parameterization, which preserves angles and hence the local geometry~\cite{gu2008computational}, can be achieved by many approaches such as harmonic energy minimization~\cite{levy2002least,gu2004genus,choi2015flash,yueh2017efficient}, geometric flow~\cite{luo2004combinatorial,jin2008discrete,crane2013robust}, conformal welding~\cite{choi2020parallelizable}, and circle patterns~\cite{kharevych2006discrete}.
However, it is well-known that conformal parameterization may result in large area distortions. By contrast, authalic parameterization focuses on area preservation and can be achieved via various methods such as Lie advection~\cite{zou2011authalic}, optimal mass transport~\cite{zhao2013area,su2016area,cui2019spherical}, density-equalizing map~\cite{choi2018density,lyu2024spherical,lyu2026ellipsoidal}, and stretch energy minimization~\cite{yueh2023theoretical}. {A Riemannian gradient descent method has also been developed for computing spherical area-preserving mappings by minimizing a stretch-energy functional~\cite{SuYu24}.} Moreover, as a generalization of conformal maps, quasi-conformal maps~\cite{ahlfors2006lectures,lehto1973quasiconformal} have been widely applied to surface parameterization~\cite{zeng2009surface,zeng2012computing,lui2013texture,choi2015fast,chien2016bounded,choi2018linear} to control the angular distortion and preserve bijectivity. Furthermore, other parameterization methods have also considered achieving a balance between angle and area distortions~\cite{fu2015computing, wang2016arap++,nadeem2016spherical, hu2017advanced,lyu2024bijective}.

To represent 3D shapes in terms of not only their surface geometries but also their interior structures, volumetric meshes are commonly used. Analogous to surface parameterization, it is natural to utilize volumetric parameterization to map 3-manifolds onto simple canonical domains for shape analysis and modelling. However, compared with surface parameterization, volumetric parameterization is generally much more challenging due to the complex internal structure of those solid manifolds. In recent years, there has been an increasing interest in volumetric parameterization methods. For instance, Wang et al.~\cite{wang2003volumetric} developed a method for computing volumetric harmonic maps. Naitsat et al.~\cite{naitsat2015volumetric}, Lee et al.~\cite{lee2016landmark}, and Zhang et al.~\cite{zhang2022unifying} proposed methods for computing volumetric quasi-conformal mappings. In~\cite{su2017volume}, Su et al. proposed an algorithm for computing volumetric optimal mass transport mappings. In~\cite{yueh2019novel,yueh2020new,yueh2021projected}, Yueh et al. developed volumetric parameterization methods via stretch energy minimization, and Tan et al.~\cite{tan2025dimensional} extended this framework into high-dimensional cases. More recently, Lyu et al.~\cite{lyu2025volumetric} developed the 3DDEQCMap method for the volumetric parameterization of simply connected 3-manifolds onto a solid ball with controlled quasi-conformal and volumetric distortions. Other recent works on volumetric parameterization include close-to-conformal volumetric mappings~\cite{paille2012conformal,chern2015close}, and stretch-distortion energy~\cite{jin2015stretch}, density-equalizing reference map~\cite{choi2021volumetric}, and isovolumetric energy minimization~\cite{liu2026volume}.

 \begin{figure}[t!]
    \centering
    \includegraphics[width=1\textwidth]{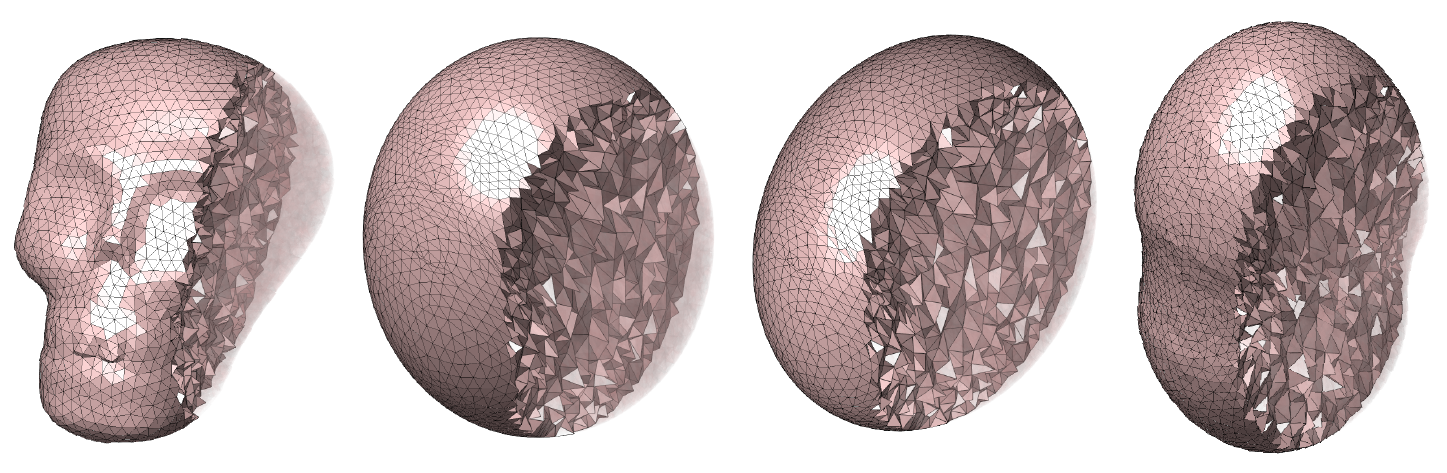}
    \caption{\textbf{An illustration of the proposed adaptive volumetric parameterization framework.} From left to right: An input 3-manifold skull model, a volume-preserving parameterization onto a prescribed ellipsoidal domain, a volume-preserving parameterization onto an optimized ellipsoidal domain, and a free-boundary volumetric parameterization based on a user-defined prescribed mass distribution.} 
    \label{fig:illustration}
\end{figure}

While many volumetric parameterization methods have been developed in the above-mentioned works, most of them have only considered the parameterization of volumetric meshes onto standardized domains such as a solid ball. For solid objects with more extreme geometries, such as highly elongated shapes, parameterizing them onto such domains may introduce significant geometric distortion, thereby posing challenges for subsequent shape processing and analysis tasks. To address this issue, in this work we propose a novel adaptive volumetric parameterization framework for simply connected 3-manifolds (see Fig.~\ref{fig:illustration} for illustrations). {Unlike 3DDEQCMap~\cite{lyu2025volumetric}, which uses a fixed solid-ball target, and the ellipsoidal density-equalizing map (EDEM) method~\cite{lyu2026ellipsoidal}, which concerns ellipsoidal surface parameterization, our model incorporates target-domain adaptation into volumetric parameterization. Specifically, we formulate a unified variational model that couples the volumetric mapping and target-domain geometry while balancing local shape and mass distortions.} 
Within this framework, we consider three different target domains, each corresponding to different geometric constraints. For the prescribed ellipsoidal domain, the shape and radii of the target ellipsoid are fixed in advance, providing a common canonical domain for volumetric parameterization. For the volume-normalized adaptive ellipsoidal domain, the ellipsoid radii are allowed to vary while the target volume is preserved, thus enabling the ellipsoid radii to adapt to the input geometry and the prescribed mass distribution. For the sea-embedded free-boundary case, the ellipsoidal constraint is further relaxed, allowing the target boundary to deform within the sea. The effectiveness of our proposed methods has been verified through experimental results and practical applications. The contributions of this paper can be summarized as follows:
\begin{itemize}
    \item We formulate a target-domain-adaptive variational model for volumetric parameterization of simply connected 3-manifolds that jointly optimizes the volumetric mapping and target-domain geometry while balancing local shape and mass distortions.

    \item We develop three volumetric parameterization models with increasing target-domain flexibility, extending the fixed-ball setting to prescribed solid ellipsoids, adaptive ellipsoids with radii optimized under a fixed-volume constraint, and free-boundary domains embedded in an auxiliary sea. This flexibility allows the target geometry to better accommodate the input shape and prescribed mass distribution, thereby providing greater control over local shape and mass distortions. All three algorithms combine 3D quasi-conformal shape updates and diffusion-based density-equalizing updates, incorporate the target-domain constraints of each model, and include a geometric correction step to remove tetrahedral foldings.

    \item We demonstrate the effectiveness of the proposed model in controlling local shape and mass distortions through numerical experiments. The proposed methods can be applied to various tasks, including 3-manifold remeshing, registration, and morphing, illustrating their practicality.
\end{itemize}

The rest of the paper is organized as follows. In Section~\ref{sec:mathematical_model}, we introduce the mathematical models for adaptive target-domain volumetric parameterization in detail. In Section~\ref{sec:proposed_alg}, we describe the numerical details of our proposed algorithms. Experimental results are presented in Section~\ref{sec:experimental_results} to assess the performance of our proposed methods. In Section~\ref{sec:applications}, we introduce the applications of our methods to 3-manifold remeshing, registration, and morphing. We conclude our work and discuss possible future directions in Section~\ref{sec:conclusion}.

\section{Mathematical model}\label{sec:mathematical_model}
In our proposed framework, we develop a variational model to compute volumetric parameterizations of simply connected 3-manifolds. Unlike previous approaches, this model simultaneously controls both local shape and mass distortions while controlling fold-over throughout the parameterization process. In the following, we introduce the variational model and design a numerical algorithm to solve it.

\subsection{Target domain adaptive variational model}
Let $\mathcal M$ be a simply connected 3-manifold. In contrast to conventional volumetric parameterization methods that rely on a predefined target domain, our approach treats the target domain $\Omega\subset\mathbb R^3$ as an intrinsic variable of the model. Specifically, we aim to find a volumetric map $f:\mathcal M \rightarrow \Omega$ as well as the parameter domain $\Omega$. {We restrict the admissible mappings to orientation-preserving, nondegenerate bijections satisfying $\det(J_f)>0$. The corresponding energy is defined as}
\begin{equation}\label{eq:combined_model}
    E(f,\Omega)=\alpha\int_{\mathcal M}|\log K_f|^2 + \beta\int_{\Omega}|\nabla \rho_{f,\Omega}|^2,
\end{equation}
where $\alpha,\beta \ge 0$ balance the local shape distortion and the mass distortion. We denote the first and second terms by $E_{\mathrm{shape}}(f)$ and
$E_{\mathrm{mass}}(f,\Omega)$, respectively. Here, $K_f$ is the 3D quasi-conformal dilation of $f$, and $\rho_{f,\Omega}$ denotes the push-forward density of the prescribed mass measure under $f$. The two energy terms will be explained in detail in the following two subsections. Note that the formulation naturally extends the 3DDEQCMap formulation in~\cite{lyu2025volumetric} by incorporating $\Omega$ as a variable, enabling us to optimize not only the volumetric mapping but also the target domain.

The proposed variational target-domain adaptive parameterization problem is given by
\begin{equation}
    \min_{\Omega\in\mathcal A,\ f:\mathcal M\to\Omega} E(f,\Omega),
\end{equation}
where $\mathcal A$ denotes a family of admissible target domains. The energy formulation also includes several special cases. By setting $\alpha = 1$ and $\beta = 0$, the model focuses on reducing local shape distortion. In contrast, when $\alpha = 0$ and $\beta = 1$, the model aims solely to equalize density. Furthermore, if $\alpha, \beta > 0$, the model can minimize the joint distortion.

The formulation in Eq.~\eqref{eq:combined_model} includes both fixed-target and target-adaptive parameterization models. When the admissible family $\mathcal A$ contains only one prescribed domain, the problem reduces to a conventional fixed-target parameterization problem. When $\mathcal A$ contains multiple admissible domains, the target geometry becomes part of the optimization variables, allowing the parameter domain to adapt together with the volumetric map.

\subsection{Local shape distortion via 3D quasi-conformality} 
We first introduce the 3D quasi-conformal dilation $K$ developed in~\cite{chen2025newgeometricrepresentation3d,lyu2025volumetric}. 
Let $f: \mathcal M \to\Omega$ be a volumetric map, and let $J_f(x)$ denote its Jacobian matrix at $x \in \mathcal M$. Following the 3D quasi-conformal representation in~\cite{chen2025newgeometricrepresentation3d}, we consider the pointwise modified polar decomposition $J_f(x)=R(x)D(x)$, where $R(x)\in SO(3)$ is a rotation matrix and $D(x)$ is a symmetric matrix with eigenvalues $\lambda_1(x)\geq\lambda_2(x)\geq\lambda_3(x)$. In the following, we omit the explicit dependence on $x$ when no ambiguity arises.
The 3D quasi-conformality is defined as $q=(\Lambda,\Theta)$, where $\Lambda=(\lambda_1,\lambda_2,\lambda_3)$ contains the eigenvalues of $D$ and $\Theta=(\theta_x,\theta_y,\theta_z)$ contains the associated Euler angles. Here, $\Lambda$ describes the principal stretches, while $\Theta$ determines their orientations. The mapping is conformal if and only if $\lambda_1=\lambda_2=\lambda_3$, or equivalently, $\lambda_1=\lambda_3$. Moreover, with a suitable choice of the rotational component in the modified polar decomposition, the sign of $\lambda_3$ coincides with that of $\det(\mathbf J_f)$~\cite{lyu2025volumetric}.

To effectively quantify the local shape distortion of the mapping, the 3D quasi-conformal dilation is introduced as follows~\cite{lyu2025volumetric}: $K(\lambda_1,\lambda_2,\lambda_3) = \begin{cases} \frac{\lambda_1}{\lambda_3}, & \text{if } \det(J_f) \neq 0, \\ \infty, & \text{if } \det(J_f)  = 0, \end{cases}$ where $\lambda_1 \geq \lambda_2 \geq\lambda_3$ are the eigenvalues of the matrix $D$. The dilation $K$ quantifies shape distortion via its magnitude and encodes orientation through its sign. Specifically, when $\infty>K>0$, the mapping is locally orientation-preserving and a value of $K$ close to $1$ indicates small local shape distortion.

\subsection{Mass distortion via density equalization}
We next introduce the density term. Let $(\mathcal M,\omega)$ be the input measured solid manifold, where $\omega$ denotes the prescribed mass measure on $\mathcal M$. For a volumetric map $f:\mathcal M\to \Omega$, the push-forward measure $f_*\omega$ describes the distribution of prescribed mass over the target domain $\Omega$.

The volumetric mapping $f$ is mass-preserving when the push-forward measure $f_*\omega$ is equivalent to the target Lebesgue measure $dV$ up to a constant scaling factor:
\begin{equation}\label{eq:push_forward_condition}
    f_*\omega = k_{\Omega} dV,
\end{equation}
where $k_\Omega = \frac{\omega(\mathcal M)}{\operatorname{Vol}(\Omega)}.$

In general, there exists a density function $\rho_{f,\Omega}:\Omega\to(0,\infty)$ such that
\begin{equation}
    f_{*}\omega = \rho_{f,\Omega} \, dV.
\end{equation}
Therefore, Eq.~\eqref{eq:push_forward_condition} is equivalent to
\begin{equation}
    \rho_{f,\Omega}(x)=k_\Omega
\end{equation}
for $x\in\Omega$. At this density-equalized state, $E_{\mathrm{mass}}(f,\Omega)=0$.

\subsection{Admissible target domains}\label{sec:admissible_domains}
The target-domain adaptive model in Eq.~\eqref{eq:combined_model} depends on the choice of the admissible family $\mathcal A$. In this work, we consider three target-domain families: a prescribed ellipsoidal domain, a volume-normalized adaptive ellipsoidal domain, and a sea-embedded free-boundary domain.

\subsubsection{Prescribed ellipsoidal domain}\label{sec:prescribed_ellipsoid}
We first consider the case where the target domain is a prescribed solid
ellipsoid. For given radii $(a^0,b^0,c^0)$, the solid ellipsoid is 
\begin{equation}
    \mathbb{E}_{a^0,b^0,c^0}=\left\{(x,y,z)\in\mathbb R^3:\frac{x^2}{(a^0)^2}+\frac{y^2}{(b^0)^2}+\frac{z^2}{(c^0)^2}\le 1\right\}.
\end{equation}
The corresponding admissible family is 
\begin{equation}
    \mathcal A_{\mathrm{fix}}=\{\mathbb{E}_{a^0,b^0,c^0}\}.
\end{equation}
Therefore, the target-domain adaptive problem reduces to
\begin{equation}\label{eq:prescribed_ellispoid}
    \min_{f:\mathcal M\to \mathbb{E}_{a^0,b^0,c^0}}
    E(f) = \alpha\int_{\mathcal M}|\log K_f|^2 + \beta\int_{\mathbb{E}_{a^0,b^0,c^0}}|\nabla\rho_{f}|^2.
\end{equation}
Here, $\rho_f$ denotes the density defined on the fixed target ellipsoid $\mathbb{E}_{a^0,b^0,c^0}$.

\subsubsection{Volume-normalized adaptive ellipsoidal domain}\label{sec:sec:adaptive_ellipsoid_model}
The above method fixes the target shape throughout the optimization. However, a prescribed ellipsoid may not match the geometric proportions of the input 3-manifold, leading to unnecessary local shape distortion. To address this issue, we introduce an adaptive ellipsoidal domain model, which allows the ellipsoidal radii to vary. 

For $a,b,c>0$, let 
\begin{equation}
    \mathbb{E}_{a,b,c}=\left\{(x,y,z)\in\mathbb R^3:\frac{x^2}{a^2}+\frac{y^2}{b^2}+\frac{z^2}{c^2}\le 1\right\}.
\end{equation}
Since 
\begin{equation}
    |\mathbb{E}_{a,b,c}|=\frac{4\pi}{3}abc,
\end{equation}
fixing the product $abc$ is equivalent to fixing the volume of the ellipsoidal target. We define the volume-normalized
adaptive ellipsoidal domains by
\begin{equation}\label{eq:radii_constraint}
    \mathcal A_{\mathrm{ellipsoidal}}
    =
    \left\{
    \mathbb{E}_{a,b,c}: abc=P_0,\quad a,b,c>0
    \right\},
\end{equation}
where $P_0>0$ is determined by the initial solid ellipsoid.

The corresponding optimization problem is
\begin{equation}\label{eq:adaptive_ellipsoid_model}
    \min_{\mathbb{E}_{a,b,c}\in\mathcal A_{\mathrm{ellipsoidal}},\ f:\mathcal M\to \mathbb{E}_{a,b,c}}
    E(f,\mathbb{E}_{a,b,c}) = \alpha\int_{\mathcal M}|\log K_f|^2 + \beta\int_{\mathbb{E}_{a,b,c}}|\nabla\rho_{f,\mathbb{E}_{a,b,c}}|^2.
\end{equation}
The constraint $abc=P_0$ removes the global scaling ambiguity and ensures that the model optimizes the aspect ratio of the target ellipsoid rather than its overall scale.

\subsubsection{Free-boundary target domain}
The adaptive ellipsoidal model still restricts the target domain to an ellipsoidal parametric family. In this section, we further relax this constraint and consider a free-boundary target model. In this free-boundary model, the target boundary is allowed to deform during the parameterization process.

In this case, the target domain is not restricted to be an ellipsoid. Instead, we consider a family of admissible target domains
\begin{equation}\label{A_free}
    \mathcal A_{\mathrm{free}}
     = 
     \left\{
    \Omega \subseteq \bar\Omega:
    \Omega\text{ is a solid domain homeomorphic to }\mathcal M
    \right\}.
\end{equation}
Here, $\bar\Omega = \mathbb{E}_{\bar a,\bar b,\bar c}$ denotes the fixed outer computational ellipsoid with radii $(\bar a,\bar b,\bar c)$. These radii are determined at initialization from the initial ellipsoidal domain and the auxiliary sea, and remain unchanged throughout the optimization.

The corresponding parameterization problem is
\begin{equation}
    \min_{\Omega\in \mathcal A_{\mathrm{free}},\ f:\mathcal M\to\Omega}
    E(f,\Omega) = \alpha\int_{\mathcal M}|\log K_f|^2 + \beta\int_{\Omega}|\nabla\rho_{f,\Omega}|^2.
\end{equation}

Compared with the ellipsoidal families, the free-boundary model relaxes the prescribed geometric constraint on the target domain and allows the boundary to adapt more freely to the input local shape and density distribution. In the numerical algorithm, this deformation is realized using an auxiliary sea region. The details of this sea-embedded realization will be introduced in Section~\ref{sec:create_sea}.

\begin{remark}
The containment constraint in~\ref{A_free} removes an unbounded scaling direction of the free-boundary problem. For $s>0$, let $f_s = sf$ and $\Omega_s = s\Omega$. Since $K_{f_s} = K_f$ and the push-forward density satisfies
\begin{equation}
    \rho_{f_s,\Omega_s}(y) =s^{-3}\rho_{f,\Omega}(y/s),
\end{equation}
a change of variables yields
\begin{equation}
    E(f_s,\Omega_s) = \alpha E_{\mathrm{shape}}(f) + \beta s^{-5}E_{\mathrm{mass}}(f,\Omega).
\end{equation}
Consequently, if $\beta E_{\mathrm{mass}}(f,\Omega)>0$, the energy decreases under uniform expansion of the target domain, while the shape energy remains unchanged. The constraint $\Omega \subseteq \bar\Omega$ excludes arbitrarily large dilations of this form.
\end{remark}

\section{Proposed algorithm}\label{sec:proposed_alg}
In this section, we propose three numerical algorithms for the target domain adaptive models introduced in Section~\ref{sec:mathematical_model}.

\subsection{Prescribed ellipsoidal model}\label{sec:fix_shape}
Consider a 3D solid manifold $\mathcal{M}$ in $\mathbb R^3$ with a closed surface boundary $\partial \mathcal{M}$ discretized as a tetrahedral mesh $(\mathcal{V}, \mathcal{E},\mathcal{T})$, where $\mathcal{V}$ is the set of vertices, $\mathcal{E}$ is the set of edges, and $\mathcal{T}$ is the set of tetrahedral elements. Our goal is to develop an energy minimization algorithm for computing a bijective volumetric parameterization of $\mathcal{M}$ onto a prescribed ellipsoid balancing the local shape distortion (based on 3D quasi-conformality) and the mass distortion (based on a user-defined mass distribution). { To achieve this, we follow the solid ball parameterization method in 3DDEQCMap~\cite{lyu2025volumetric} and extend the computation to a prescribed solid ellipsoid. When $a=b=c=1$, the prescribed ellipsoidal model reduces exactly to the 3DDEQCMap~\cite{lyu2025volumetric}, both in its mathematical formulation and numerical implementation.}

\subsubsection{Initial solid ellipsoid}\label{sec:initial_mapping}
Given a prescribed ellipsoid $\mathbb E_{a,b,c}$, we first compute an initial volumetric map $f^0:\mathcal{M}\rightarrow \mathbb{E}_{a,b,c}$. The initial map is computed in two steps. Firstly, we compute the boundary ellipsoidal parameterization $f^0_{\partial}:\partial\mathcal M\to\partial\mathbb E_{a,b,c}$ utilizing the ellipsoidal density-equalizing map (EDEM) method~\cite{lyu2026ellipsoidal}. Then, the interior mapping is obtained by solving the Laplace equation
\begin{equation}\label{eq:Laplace}
    \left \{
    \begin{array}{ll}
    \Delta f^0 = 0     & \text{in} \quad \mathcal{M} \setminus \partial \mathcal{M}, \\
    f^0 \left(\partial \mathcal{M} \right) = \partial \mathbb{E}_{a,b,c}.    & 
    \end{array}    
    \right.
\end{equation}
The resulting map is used as the initial map for the subsequent balanced optimization.

\subsubsection{Local-shape descent}\label{sec:optimize_shape}
The local-shape descent follows the 3DDEQCMap procedure in~\cite{lyu2025volumetric}. Let $f^n$ be the current ellipsoidal map with 3D quasi-conformality $q^n=(\Lambda^n,\Theta^n)$, where $\Lambda^n=(\lambda_1^n,\lambda_2^n,\lambda_3^n)$. As described in~\cite{lyu2025volumetric}, one can update the 3D quasi-conformality as $q^{n+1} = (\Lambda^{n+1}, \Theta^{n+1})$, where $\Theta^{n+1} = \Theta^{n}$ and $\Lambda^{n+1} =
    \bigl(
    \lambda_1^n-\tau^nR^n,\,
    \lambda_2^n,\,
    \lambda_3^n+\tau^nL^n
    \bigr)$, where $R^n=\lambda_1^n-\lambda_2^n$,  $L^n=\lambda_2^n-\lambda_3^n$, $\tau^n=\frac{K^n-1}{(K^n-1)+\mathbf C}$, $K^n=\frac{\lambda_1^n}{\lambda_3^n}$, and $\mathbf C>0$. If
$\lambda_1^n\geq\lambda_2^n\geq\lambda_3^n>0$, this update preserves the eigenvalue ordering and satisfies $K^{n+1}\leq K^n$. Consequently, the shape energy will be reduced. Readers are referred to~\cite{lyu2025volumetric} for more details.

\subsubsection{Density-equalizing descent}\label{sec:optimize_mass}
We next compute a density-equalizing descent direction for the mass term $E_{\text{mass}}$. The density-equalizing step is motivated by the diffusion equation~\cite{gastner2004diffusion}
\begin{equation}\label{eq:diffusion}
    \partial_t \rho = \Delta \rho,
\end{equation}
which drives the density from high-density regions to low-density regions and therefore reduces spatial density variation. By diffusion theory and Fick's law, the velocity field is given by
\begin{equation}\label{eq:velocity_field}
    \mathbf{v}(\mathbf{x}(t),t) = -\frac{\nabla \rho(\mathbf{x}(t),t)}{\rho(\mathbf{x}(t),t)}.
\end{equation}
The tracer of $\mathbf{x}$ at time $t$ can be computed by 
\begin{equation}\label{eq:integration_tracer}
    \mathbf{x}(t) = \mathbf{x}(0) + \int^{t}_0 \mathbf{v}(\mathbf{x}(\tau),\tau) d\tau.
\end{equation}
As $t\rightarrow \infty$, the density induced by the diffusion process converges to a constant, yielding the density-equalizing map.

\begin{remark}
On a fixed ellipsoidal domain $\Omega=\mathbb E_{a,b,c}$ with Neumann boundary condition, the diffusion equation $\partial_t\rho=\Delta\rho$ dissipates the density Dirichlet energy:
\begin{equation}
    \frac{d}{dt}\int_{\Omega}|\nabla\rho|^2\,dV
    =
    -2\int_{\Omega}|\Delta\rho|^2\,dV
    \leq 0.
\end{equation}
Thus, the diffusion process decreases $E_{\mathrm{mass}}$ on a fixed
target domain.
\end{remark}

Let $f^n:\mathcal M \to \mathbb{E}_{a,b,c}$ be the $n$-th iteration mapping with density $\rho^n$. In the discrete case, we solve the density equalization and vertex update problem in the ellipsoidal domain discretized in the form of tetrahedral meshes, and the solving procedure follows from prior numerical schemes as detailed in~\cite{choi2018density,lyu2025volumetric}. Specifically, the density on each tetrahedral element is defined by:
\begin{equation}
    \rho^{\mathcal{T}}_n = \frac{m_T}{\operatorname{Vol}(T)},
\end{equation}
where $m_T$ is the prescribed mass assigned to $T$.

Then, we utilize the semidiscrete backward Euler method to solve the diffusion equation~\eqref{eq:diffusion}:
\begin{equation}\label{eqt:dis_diffusion}
    \frac{\rho^{\mathcal{V}}_{n+1} - \rho^{\mathcal{V}}_{n}}{\delta t} = \Delta_{n}\rho^{\mathcal{V}}_{n+1},
\end{equation}
where $\rho^{\mathcal{V}}_{n}$ is the $n$-th iterative vertex density, $\delta t$ is the timestep, and $\Delta_{n}$ is the Laplace--Beltrami operator. On the tetrahedral mesh, the Laplace--Beltrami operator can be written as:
\begin{equation}
    \Delta_{n} = -A^{-1}_nL_n.
\end{equation}
Here $A_n$ is a $|\mathcal{V}| \times |\mathcal{V}|$ diagonal matrix such that
\begin{equation}\label{eq:A_n}
        \left(A_n\right)_{i} = \frac{1}{4} \sum_{\mathcal{T}_i \in \mathcal{N}^{\mathcal{T}}(\mathbf{x}_i)} \text{Vol}\left( \mathcal{T}_i \right),
\end{equation}
and $L_n$ is the volumetric cotangent Laplacian matrix defined by
\begin{equation}\label{eq:L_n}
    \left(L_n\right)_{ij} = 
      \left \{
    \begin{array}{ll}
     h_{i,j}    & \text{if }  \left[\mathbf{x}_i,\mathbf{x}_j \right] \in \mathcal{E}, \\
     -\sum_{k\neq i} h_{i,k}    & \text{if }  j = i, \\
     0      & \text{otherwise}. \\
    \end{array}  
    \right.
\end{equation}
Here, $h_{i,j} = -\sum_{s = 1}^{N} l_{s} \frac{\cot \left(\theta_{s}\right)}{12}$ is the cotangent weight for volumetric Laplacian~\cite{wang2003volumetric}, where $N$ is the number of tetrahedra sharing the edge $[\mathbf{x}_i,\mathbf{x}_j]$. $\theta_s$ is the dihedral angle opposite to the edge in the $s$-th incident tetrahedron, and $l_s$ is the length of the corresponding opposite edge.

Therefore, Eq.~\eqref{eqt:dis_diffusion} can be written as
\begin{equation}\label{eq:matrix_diffusion}
    \rho^{\mathcal{V}}_{n+1} = \left(A_n + \delta t L_n \right)^{-1} \left( A_n \rho^{\mathcal{V}}_{n} \right).
\end{equation}
By solving this linear system, we can obtain the vertex density $\rho^{\mathcal{V}}_{{n+1}}$.

Next, we compute the density gradient on each tetrahedral element as in~\cite{lyu2025volumetric}. For a tetrahedron $T=[\mathbf{x}_i,\mathbf{x}_j,\mathbf{x}_k,\mathbf{x}_l]$, the vertex density is linearly interpolated via the Whitney 0-form:
\begin{equation}
    \rho^{\mathcal{V}}_{{n+1}}(\mathbf{x}) = \rho^{\mathcal{V}}_{{n+1}}(\mathbf{x}_i)\Gamma^W_i(\mathbf{x}) + \rho^{\mathcal{V}}_{{n+1}}(\mathbf{x}_j)\Gamma^W_j(\mathbf{x}) + \rho^{\mathcal{V}}_{{n+1}}(\mathbf{x}_k)\Gamma^W_k(\mathbf{x}) + \rho^{\mathcal{V}}_{{n+1}}(\mathbf{x}_l)\Gamma^W_l(\mathbf{x}),
\end{equation}
where $\Gamma^W_s(\mathbf{x})$ denotes the linear hat function associated with vertex $\mathbf{x}_s$. Since $\nabla \Gamma^W_i = -\frac{S_i}{6\text{Vol}(\mathcal{T})}$, the density gradient $\nabla \rho^{\mathcal{T}}_{n+1}(\mathcal{T})$ is given by
\begin{equation}
        \nabla \rho_{{n+1}}^{\mathcal{T}}(\mathcal{T}) = -\frac{\rho^{\mathcal{V}}_{{n+1}}\left(\mathbf{x}_i\right) S_i + \rho^{\mathcal{V}}_{{n+1}}\left(\mathbf{x}_j\right) S_j + \rho^{\mathcal{V}}_{{n+1}}\left(\mathbf{x}_k\right) S_k + \rho^{\mathcal{V}}_{{n+1}}\left(\mathbf{x}_l\right) S_l}{6 \ \text{Vol}(\mathcal{T})},
\end{equation}
where the edges are denoted as $\vec{e}_{ij} = {[\mathbf{x}_i,\mathbf{x}_j]}$, and
$S_i = (\vec{e}_{kl}\times \vec{e}_{kj}),S_j = (\vec{e}_{il}\times \vec{e}_{ik}), S_k = (\vec{e}_{ij}\times \vec{e}_{il}), S_l = (\vec{e}_{ik}\times \vec{e}_{ij})$. We can then easily convert $\nabla \rho_{n+1}^{\mathcal{V}}$ from $\nabla \rho_{n+1}^{\mathcal{T}}$ via the above-mentioned tetrahedral-to-vertex conversion matrix $M^{\mathcal{T}}_{\mathcal{V}}$:
\begin{equation}
    \nabla \rho_{{n+1}}^{\mathcal{V}}(\mathbf{x}_i) = \frac{\sum_{T\in N_T(\mathbf{x}_i)} \operatorname{Vol}(T)\nabla \rho_{{n+1}}^{\mathcal{T}}}{\sum_{T\in N_T(\mathbf{x}_i)} \operatorname{Vol}(T)}.
\end{equation}

The velocity field on each vertex is given as follows:
\begin{equation}\label{eq:vertex_velocity}
    \mathbf{v}^{\mathcal{V}}_{{n+1}}(\mathbf{x}_i) =- \frac{\nabla \rho_{{n+1}}^{\mathcal{V}}\left(\mathbf{x}_i\right)}{\rho_{{n+1}}^{\mathcal{V}}\left(\mathbf{x}_i\right)}.
\end{equation}
It is noteworthy that our goal is to compute an ellipsoidal volumetric parameterization. However, in the discrete case, the velocity on the boundary may not lie in the tangent space. To address this issue, we follow the idea of the EDEM method~\cite{lyu2026ellipsoidal} and project the boundary velocity onto the tangent plane of the ellipsoidal boundary. Note that the equation of the ellipsoidal boundary is 
\begin{equation}
    \frac{x^2}{a^2} + \frac{y^2}{b^2} + \frac{z^2}{c^2} = 1,
\end{equation}
and the outward unit normal vector at $\mathbf{x}_{\text{bdy}}$ is 
\begin{equation}
\mathbf{n}^{\mathcal{V}}(\mathbf{x}_{\text{bdy}}) = \frac{\left(\frac{x}{a^2}, \frac{y}{b^2}, \frac{z}{c^2} \right)}{\left( \left(\frac{x}{a^2} \right)^2 + \left(\frac{y}{b^2} \right)^2 + \left(\frac{z}{c^2} \right)^2 \right)^{\frac{1}{2}}}.
\end{equation}
The projected boundary velocity can be computed via:
\begin{equation} \label{eqt:ellipsoidalprojection}
    \widetilde{\mathbf{v}}^{\mathcal{V}}_{{n+1}}(\mathbf{x}_{\text{bdy}}) = \mathbf{v}^{\mathcal{V}}_{{n+1}}(\mathbf{x}_{\text{bdy}}) - \left( \mathbf{v}^{\mathcal{V}}_{{n+1}}(\mathbf{x}_{\text{bdy}}) \cdot \mathbf{n}^{\mathcal{V}}(\mathbf{x}_{\text{bdy}}) \right) \mathbf{n}^{\mathcal{V}}(\mathbf{x}_{\text{bdy}}),
\end{equation}
where $\mathbf{x}_{\text{bdy}}$ denotes the boundary vertex. Ultimately, the velocity field $\overline{\mathbf{v}}^{\mathcal{V}}_{{n+1}}$ of diffusion flow is given by
\begin{equation}\label{eq:velocity_final}
        \overline{\mathbf{v}}^{\mathcal{V}}_{{n+1}}(\mathbf{x}_i) = 
      \left \{
    \begin{array}{ll}
        \widetilde{\mathbf{v}}^{\mathcal{V}}_{{n+1}}(\mathbf{x}_i)  & \text{if } \mathbf{x}_i \in \partial \mathbb E_{a,b,c},\\
    \mathbf{v}^{\mathcal{V}}_{{n+1}}(\mathbf{x}_i)     & \text{otherwise}, 
    \end{array}    
    \right.
\end{equation}

Utilizing the above formulas, we can update the positions of all vertices based on Eq.~\eqref{eq:integration_tracer}:
\begin{equation}\label{eq:update_f}
    \mathbf{x}_i(t_{n+1}) = \mathbf{x}_i(t_{n}) { +} \delta t \ \overline{\mathbf{v}}^{\mathcal{V}}_{{n+1}}(\mathbf{x}_i),
\end{equation}
where $\delta t$ is the time-step size. While the projection step can eliminate the normal components of the boundary velocity, there may still be some numerical errors that cause the boundary vertices to move out of the ellipsoidal boundary surface. To enforce the boundary constraint, we divide the coordinates of each boundary vertex $\mathbf{x}_{\text{bdy}}(t_{n+1}) = (x,y,z)$ by $\sqrt{\frac{x^2}{a^2} + \frac{y^2}{b^2} + \frac{z^2}{c^2}}$. The descent process of the mass energy  $E_{\text{mass}}$ for ellipsoidal volumetric mappings is summarized in Algorithm~\ref{alg:optimize_E2}.

\begin{algorithm}[h]
\caption{Descent of the mass energy  $E_{\text{mass}}$ for ellipsoidal volumetric mappings}
\label{alg:optimize_E2}

\KwIn{A simply connected 3-manifold $\mathcal{M}$, a user-defined prescribed mass distribution, the ellipsoidal radii $(a,b,c)$, and a solid ellipsoid map $f^n:\mathcal{M}\to\mathbb{E}_{a,b,c}$.
}

\KwOut{An updated solid ellipsoid map
$f^{n+1}:\mathcal{M}\to\mathbb{E}_{a,b,c}$.
}

Compute the density $\rho_n^{\mathcal{T}}$ on the current volumetric ellipsoidal mapping result $f^n(\mathcal{M})$\;

Set $\mathbf{x}_i(t_n)=f^n(v_i)$ for every vertex $v_i\in\mathcal{V}$\;

Compute $A_n$ and $L_n$ using Eqs.~\eqref{eq:A_n} and~\eqref{eq:L_n}\;

Obtain $\rho_{n+1}^{\mathcal{V}}$ by solving the diffusion equation~\eqref{eq:matrix_diffusion}\;

Compute the velocity field and perform the ellipsoidal projection to obtain $\overline{\mathbf{v}}_{n+1}^{\mathcal{V}}$  using Eqs.~\eqref{eq:vertex_velocity}--\eqref{eq:velocity_final}\;

Update the volumetric ellipsoidal mapping $f^{n+1}=\mathbf{x}(t_{n+1})$ using Eq.~\eqref{eq:update_f}\;

\end{algorithm}

\subsubsection{Coupled update}
After obtaining the descent directions of $E_{\text{shape}}$ and $E_{\text{mass}}$, we now combine the local shape and density updates in the 3D quasi-conformality field following the approach in~\cite{lyu2025volumetric}. Let $f^n:\mathcal M\to E_{a,b,c}$ be the volumetric ellipsoidal mapping at the $n$-th iteration with corresponding 3D quasi-conformality $q^n = (\Lambda^n, \Theta^n)$. 

For the local shape term $E_{\text{shape}} = \int_{\mathcal M} |\log{K} |^2$, we utilize the residual method in Section~\ref{sec:optimize_shape} following~\cite{lyu2025volumetric} to compute the updated 3D quasi-conformality $\widetilde{q}^{n+1}$. The descent direction for the shape term is given by $dq^n_1 = \widetilde{q}^{n+1} - q^n$.

Then, we transform the descent direction of the mass energy $E_{\text{mass}}$ from the physical spatial coordinates to the 3D quasi-conformality field as in~\cite{lyu2025volumetric}. Specifically, using the diffusion process in Section~\ref{sec:optimize_mass}, we can obtain the auxiliary updated volumetric ellipsoidal mapping $\hat{f}^{n+1}:\mathcal{M} \rightarrow \mathbb{E}_{a,b,c}$ with corresponding 3D quasi-conformality $\hat{q}^{n+1}$. Hence, $dq^n_2$ can be obtained by $dq^n_2 = \hat{q}^{n+1} - q^n$.
Here, $\hat{f}^{n+1}$ is used to extract the density-induced variation in the quasi-conformality field.

The two updates are combined by $dq^n = \alpha dq^n_1 + \beta dq^n_2$, and the updated 3D quasi-conformality field is given by
\begin{equation}\label{eq:updated_q}
   \bar{q}^{n+1} = q^n + \delta t dq^n,
\end{equation}
where $\delta t$ is the timestep. The corresponding updated 3D quasi-conformal dilation is evaluated as $\bar{K}^{n+1} = \frac{\bar{\lambda_1}^{n+1}}{\bar{\lambda_3}^{n+1}}$, where $\bar{\lambda_1}^{n+1} = \max(\bar{\Lambda}^{n+1})$ and $\bar{\lambda_3}^{n+1} = \min(\bar{\Lambda}^{n+1})$. {We compute the update directions for the shape and mass terms separately, following Sections~\ref{sec:optimize_shape} and~\ref{sec:optimize_mass}, and combine them in Eq.~\eqref{eq:updated_q}. The resulting update, including mapping reconstruction, boundary projection, and geometric correction, does not guarantee a decrease in the complete energy at every iteration. Nevertheless, the numerical results in Section~\ref{sec:comparisom_result} show an overall decrease in the total energy across the tested examples, despite occasional local increases (see Fig.~\ref{fig:example_energy}).
}

\subsubsection{Geometric correction scheme}\label{sec:geometry_correctopn_scheme}
While the aforementioned procedure can efficiently update the 3D quasi-conformality field and optimize the energy, the bijectivity of the resulting volumetric mapping is not guaranteed. To resolve this issue, we develop the geometric correction scheme for ellipsoidal volumetric parameterization following the idea in~\cite{lyu2025volumetric} to rectify the overlaps via the 3D quasi-conformality field in this section. Let $\mathbb{E}^n_{a,b,c} = f^n(\mathcal{M})$ be the solid ellipsoid at the $n$-th iteration. Analogous to the geometric correction scheme in~\cite{lyu2025volumetric}, we handle the foldings at two stages: We first resolve the foldings on the ellipsoidal boundary, followed by resolving the foldings at the interior of the domain.

Specifically, to eliminate boundary foldings on $\partial \mathbb{E}^n_{a,b,c}$, we apply the quasi-conformal truncation in the EDEM method~\cite{lyu2026ellipsoidal} to enforce $\|\mu\|_{\infty}<1$, and map the corrected boundary back. Having obtained a folding-free boundary, we then rectify the foldings in the tetrahedral mesh $\mathbb{E}^n_{a,b,c}$ using the approach in~\cite{lyu2025volumetric}, which involves modifying the 3D quasi-conformality and reconstructing an associated 3D quasi-conformal map by solving a sparse linear system with some prescribed boundary conditions. In particular, here we can utilize the corrected boundary map at the first stage as the prescribed boundary conditions. {In our numerical algorithm, $K$ is computed from the Jacobian matrices of the reconstructed tetrahedral mesh. If the mesh contains degenerate or inverted elements, we apply the geometric correction scheme until obtaining a folding-free mesh. Then, we compute the energy on the updated corrected mesh.} Readers are referred to~\cite{lyu2025volumetric} for details.

\subsubsection{Summary}
Altogether, the proposed prescribed ellipsoidal volumetric parameterization algorithm is summarized in Algorithm~\ref{alg:unified_optimize}. In practice, we set the timestep $\delta t = 0.1$, the stopping parameter $\epsilon = 10^{-3}$, and the maximum number of {iterations} $n_{\max} = 100$.

\begin{algorithm}[h]
\caption{Prescribed ellipsoidal volumetric map}
\label{alg:unified_optimize}

\KwIn{A simply connected 3-manifold $\mathcal{M}$, a user-defined mass distribution, weighting parameters $\alpha$ and $\beta$, the stopping parameter $\epsilon$, maximum number of iterations allowed $n_{\max}$, and the threshold $K_T$.
}

\KwOut{A bijective solid ellipsoid map $f:\mathcal{M}\to\mathbb{E}_{a,b,c}$.}

Compute the initial solid ellipsoid map
$f^0:\mathcal{M}\to\mathbb{E}_{a,b,c}$\;

Compute the initial density $\rho_0^{\mathcal{T}}$, the initial 3D quasi-conformality $q_0$, and the initial 3D quasi-conformal dilation $K^0$ on $f^0(\mathcal{M})$\;

Set $n=0$;

\Repeat{$\lVert f^n-f^{n-1}\rVert<\epsilon$ \ or \ $n\geq n_{\max}$}{
Obtain the descent directions $dq_1^n$, $dq_2^n$ following Section~\ref{sec:optimize_shape} and Section~\ref{sec:optimize_mass}\;

Compute the updated $\bar{q}^{n+1}$ and $\bar{K}^{n+1}$ using the prescribed weights $\alpha, \beta$ and Eq.~\eqref{eq:updated_q}\;

Apply the correction scheme in Section~\ref{sec:geometry_correctopn_scheme} with the threshold $K_T$ to obtain the updated map $f^{n+1}$\;

Update $\rho_{n+1}^{\mathcal{T}}$ based on $f^{n+1}$\;

Set $n=n+1$\;
}

The resulting prescribed ellipsoidal volumetric map is given by $f=f^n$\;

\end{algorithm}

\subsection{Volume-normalized adaptive ellipsoidal model}\label{sec:radii_optimize}

According to Section~\ref{sec:sec:adaptive_ellipsoid_model}, we next introduce the volume-normalized adaptive ellipsoidal model, which allows the ellipsoidal target region to adaptively adjust during the optimization process.

Based on Eq.~\eqref{eq:radii_constraint}, the radii in this model must satisfy $abc=P_0$, where $P_0=a_0b_0c_0$ is determined by the initial target ellipsoid. Therefore, the target-shape update changes the aspect ratio of the ellipsoid while keeping its volume factor fixed.

\subsubsection{Decoupling the combined energy}
To solve the volume-normalized adaptive ellipsoidal model~\eqref{eq:adaptive_ellipsoid_model}, we decouple the optimization into two subproblems, where the mapping $f$ and the radii $(a,b,c)$ are optimized independently:
\begin{itemize}
\item Subproblem 1 (Mapping Update): With the ellipsoidal radii $(a,b,c)$ fixed, we update the volumetric mapping $f$ by minimizing:
\begin{equation}
    E_1(f) = \alpha \int_{\mathcal M} |\log{K}|^2+ \beta \int_{\mathbb{E}_{a,b,c}} |\nabla \rho|^2. 
\end{equation}  

\item Subproblem 2 (Radii Optimization): we perform a local ellipsoidal radii search under the volume constraint:
\begin{equation}
    E_2(a,b,c) = \alpha \int_{\mathcal M} |\log{K}|^2 + \beta \int_{\mathbb{E}_{a,b,c}} |\nabla \rho|^2,
\end{equation}
with the volume-normalization constraint $abc=P_0$.
\end{itemize}

In our numerical algorithm, these two subproblems are solved at an asymmetric iteration frequency. Specifically, subproblem 1 is executed multiple times to fully deform the mesh vertices, followed by a single execution of subproblem 2 to adjust the target domain. For fixed radii, the mapping update is exactly the prescribed ellipsoidal procedure described in Section~\ref{sec:fix_shape}. Specifically, we compute the local shape and density updates, combine them in the 3D quasi-conformality field, and apply the geometric correction before reconstructing the updated map. The target-shape update is described in the next subsection.

\subsubsection{Radii optimization}\label{sec:optimize_E_2}
For Subproblem $E_2$, we optimize the ellipsoidal radii under a volume-normalization constraint. Let $P_0 = a_0b_0c_0$ be the prescribed volume factor and $\gamma = P_0^{1/3}$. We parameterize the ellipsoidal radii by two log-radii variables $(u,v)$:
\begin{equation}
    a = \gamma e^u,\qquad b = \gamma e^v,\qquad c = \gamma e^{-u-v},
\end{equation}
which satisfies $abc = \gamma^3 e^u e^v e^{-u-v} = P_0$. Hence, the target-shape update is reduced to a two-dimensional optimization in the $(u,v)$-plane, and only the aspect ratio of the ellipsoid is changed.

For the volumetric map $f^n:\mathcal M\to\mathbb E_{a^n,b^n,c^n}$, log-radii variables are
\begin{equation}
    u^n = \log\frac{a^n}{\gamma}, \qquad v^n = \log\frac{b^n}{\gamma}.
\end{equation}
With the search step size $\delta_r$, we build a $3\times 3$ stencil in the $(u,v)$-plane:
\begin{equation}
    (\bar{u},\bar{v})=(u^n+\ell_u\delta_r,\; v^n+\ell_v\delta_r),\qquad \ell_u,\ell_v\in\{-1,0,1\}.
\end{equation}
For each $(\bar{u},\bar{v})$, the corresponding radii are 
\begin{equation}
(\bar a,\bar b,\bar c) = (\gamma e^{\bar{u}}, \gamma e^{\bar{v}}, \gamma e^{-\bar{u}-\bar{v}})
\end{equation}
such that $\bar a \bar b \bar c=P_0$.

According to the updated radii $(\bar a,\bar b,\bar c)$, we can compute the candidate ellipsoidal boundary condition 
\begin{equation}
    (x_1,x_2,x_3)\mapsto\left(\frac{\bar a}{a^n}x_1,\,\frac{\bar b}{b^n}x_2,\,\frac{\bar c}{c^n}x_3\right).
\end{equation}
Then we obtain the candidate volumetric mappings $f_{\bar{u},\bar{v}}$ with associated 3D quasi-conformality $q^n$ using the 3DQCS method in~\cite{chen2025newgeometricrepresentation3d}.

For each candidate $(\bar{u},\bar{v})$, we evaluate the full energy $E(\bar{u},\bar{v}) = E\bigl(f_{\bar{u},\bar{v}},\mathbb E_{\bar a,\bar b,\bar c}\bigr)$. Then we choose the smallest one
\begin{equation}
    (u^{n+1},v^{n+1}) = \underset{(\bar{u},\bar{v})\in\mathcal S(u^n,v^n)}{\text{argmin}} E(\bar{u},\bar{v}),
\end{equation}
where $\mathcal S(u^n,v^n)$ denotes the above $3\times 3$ set. The updated radii are 
\begin{equation}
    (a^{n+1},b^{n+1},c^{n+1}) = (\gamma e^{u^{n+1}},\gamma e^{v^{n+1}}, \gamma e^{-u^{n+1}-v^{n+1}})
\end{equation}
such that $a^{n+1}b^{n+1}c^{n+1}=P_0$.
 
Finally, we apply a uniform normalization to eliminate numerical errors and enforce the target volume $P_0$. Let $s^{n+1}=\left(\frac{a^{n+1}b^{n+1}c^{n+1}}{P_0}\right)^{1/3}.$ The mapping is updated as
\begin{equation}
(a^{n+1},b^{n+1},c^{n+1},f^{n+1})
    \leftarrow
    \left(
    \frac{a^{n+1}}{s^{n+1}},
    \frac{b^{n+1}}{s^{n+1}},
    \frac{c^{n+1}}{s^{n+1}},
    \frac{1}{s^{n+1}}f^{n+1}
    \right).
\end{equation}
This normalization uniformly rescales the map and radii, preserving the aspect ratio while satisfying the volume constraint.

The algorithm for shape optimization is summarized in Algorithm~\ref{alg:optimize_shape}.

\begin{algorithm}[h!]
\caption{Shape update of the ellipsoidal domain}
\label{alg:optimize_shape}

\KwIn{The current volumetric map
$f^n:\mathcal{M}\to\mathbb{E}_{a^n,b^n,c^n}$,
the current radii $(a^n,b^n,c^n)$, a prescribed mass distribution,
the current 3D quasi-conformality field $q^n$, and the step size $\delta_r$.}

\KwOut{The updated map $f^{n+1}$ and radii
$(a^{n+1},b^{n+1},c^{n+1})$ satisfying
$a^{n+1}b^{n+1}c^{n+1}=P_0$.}

Set $\gamma=P_0^{1/3}$\;

Compute the current log-radii variables $u^n=\log\frac{a^n}{\gamma}$, 
$v^n=\log\frac{b^n}{\gamma}$\;

Generate the local candidate set $\mathcal{S}(u^n,v^n)
=
\left\{
\left(u^n+\ell_u\delta_r,\,
v^n+\ell_v\delta_r\right)
:
\ell_u,\ell_v\in\{-1,0,1\}
\right\}$\;

\For{$(\bar{u},\bar{v})\in\mathcal{S}(u^n,v^n)$}{

    Compute the candidate radii $(\bar{a},\bar{b},\bar{c})
    =
    \left(
    \gamma e^{\bar{u}},
    \gamma e^{\bar{v}},
    \gamma e^{-\bar{u}-\bar{v}}
    \right)$\;

    Rescale the boundary image from
    $\partial\mathbb{E}_{a^n,b^n,c^n}$ to
    $\partial\mathbb{E}_{\bar{a},\bar{b},\bar{c}}$\;

    Reconstruct the candidate map $f_{\bar{u},\bar{v}}$ using $q^n$ and the updated boundary condition\;

    Evaluate the energy $E(\bar{u},\bar{v})
    =
    E\left(
    f_{\bar{u},\bar{v}},
    \mathbb{E}_{\bar{a},\bar{b},\bar{c}}
    \right)$\;
}

Choose $(u^{n+1},v^{n+1})
=
\underset{(\bar{u},\bar{v})\in\mathcal{S}(u^n,v^n)}
{\arg\min}
\,E(\bar{u},\bar{v})$\;

Set $(a^{n+1},b^{n+1},c^{n+1})
=
\left(
\gamma e^{u^{n+1}},
\gamma e^{v^{n+1}},
\gamma e^{-u^{n+1}-v^{n+1}}
\right)$\;

Set
$f^{n+1}=f_{u^{n+1},v^{n+1}}$\;

Apply a final uniform normalization to
$(a^{n+1},b^{n+1},c^{n+1},f^{n+1})$
to remove numerical errors\;

The resulting map and ellipsoidal radii are given by
$f^{n+1}$ and $(a^{n+1},b^{n+1},c^{n+1})$\;

\end{algorithm}

\subsubsection{Summary}

Combining the two subproblems yields the volume-normalized adaptive ellipsoidal parameterization algorithm. Starting from an initial map with radii $(a_0,b_0,c_0)$, we apply the mapping-update procedure for $H$ iterations on the current ellipsoid. The geometric correction is employed at each iteration. We then perform one target-shape update under the fixed-volume constraint. {Let $\delta_r$ denote the base log-radii search step. At the $m$-th target-shape update, we use the step size $\delta_r^{(m)}=0.9^m\delta_r$.}
The iterative process is repeated until $\|f^{n+1}-f^n\|<\epsilon$ or $n=n_{\max}$, as summarized in Algorithm~\ref{alg:shape_optimized_optimize}. In practice, the initial radii can be selected arbitrarily, the timestep is $0.1$, the initial radius step sizes are $\delta_r = 0.1$, the number of iterations for each set of fixed radii is $H = 5$, the error threshold is $\epsilon = 10^{-3}$, and the maximum number of iterations is $n_{\max} = 100$.

\begin{algorithm}[h]
\caption{Volume-normalized adaptive ellipsoidal map}
\label{alg:shape_optimized_optimize}

\KwIn{A simply connected 3-manifold $\mathcal{M}$, a user-defined mass distribution, weight parameters $\alpha$ and $\beta$, the threshold $K_T$, the initial ellipsoidal radii $(a_0,b_0,c_0)$, the step size $\delta t$, the radius step size $\delta_r$, the number of iterations $H$ performed for each fixed set of radii, the stopping parameter $\epsilon$, and the maximum number of iterations allowed $n_{\max}$.}
\KwOut{A volumetric ellipsoidal parameterization
$f:\mathcal{M}\to\mathbb{E}_{a,b,c}$ with adapted radii $(a,b,c)$.}

Set $(a,b,c)=(a_0,b_0,c_0)$\;

Compute the initial solid ellipsoid map
$f^0:\mathcal{M}\to\mathbb{E}_{a,b,c}$\;

Compute the initial density $\rho_0^{\mathcal{T}}$, the initial 3D quasi-conformality $q^0$, and the initial 3D quasi-conformal dilation $K^0$ on $f^0(\mathcal{M})$\;

Set $n=0$\;

\Repeat{$\lVert f^n-f^{n-1}\rVert<\epsilon$
\ or \ $n\geq n_{\max}$}{

Apply the descent method in Section~\ref{sec:optimize_shape} and Section~\ref{sec:optimize_mass} to compute the combined descent direction $dq^n$\;

Update the 3D quasi-conformality by $\bar{q}^{n+1}=q^n+\delta t\,dq^n$\;

Apply the correction scheme in Section~\ref{sec:geometry_correctopn_scheme} to obtain the updated map $f^{n+1}$\;

Update $\rho_{n+1}^{\mathcal{T}}$ based on $f^{n+1}$\;

\If{$n=Hm$ for some positive integer $m$}{

    Apply the shape update method (Algorithm~\ref{alg:optimize_shape}) with radius step size
    $0.9^m\delta_r$ to obtain the updated map $f^{n+1}$ and updated radii $(a,b,c)$\;

    Update $\rho_{n+1}^{\mathcal{T}}$, $q^{n+1}$, and $K^{n+1}$ on the updated ellipsoid $\mathbb{E}_{a,b,c}$\;
}

Set $n=n+1$\;

}

The resulting volume-normalized adaptive ellipsoidal map is given by $f:\mathcal{M}\to\mathbb{E}_{a,b,c}$, where $(a,b,c)$ are the adapted ellipsoidal radii\;

\end{algorithm}

\subsection{Sea-embedded free-boundary model}\label{sec:free_bdy}
Both the prescribed and volume-normalized adaptive models restrict the target domain to a strict solid ellipsoid. To further relax this target-domain constraint, we introduce a sea-embedded free-boundary target model. By embedding the target domain $\Omega$ inside an auxiliary sea domain, $\Omega$ can deform freely, which simultaneously prevents unbounded computational expansion.

\subsubsection{Construction of the sea-embedded domain}\label{sec:create_sea}
Let $f: \mathcal{M} \to \Omega$ be the initial map, where $\Omega = \mathbb{E}_{a,b,c}$ denotes a solid ellipsoidal tetrahedral mesh. Specifically, we denote the set of boundary triangular faces and the corresponding boundary vertices as $\mathcal{F}_{\partial \Omega}$ and $\mathcal{V}_{\partial \Omega}$, respectively.

The auxiliary sea region is constructed by generating several concentric layers outward along the boundary surface $\mathcal{F}_{\partial \Omega}$. Let $\bar l$ denote the mean boundary edge length in $\Omega$. The thickness of the layer is defined as
\begin{equation}
    h = \kappa\bar l,
\end{equation}
where $\kappa$ is a positive constant. For each boundary vertex $v_i\in\mathcal V_{\partial}$, we then generate the corresponding sea vertices by radial scaling:
\begin{equation}
    w_{k,i} = s_k v_i, \qquad k=1,\ldots,L,
\end{equation}
where $s_k=1+kh$ and $L$ is the number of layers.

Each boundary triangle $f=(i,j,\ell)\in \mathcal{F}_{\partial \Omega^0}$ forms a triangular prism between adjacent layers. We subdivide each prism into three tetrahedrons across all boundary faces, and all sea layers yield a tetrahedral shell mesh $\Omega_{\mathrm{sea}}$. The sea-embedded domain is then defined as
\begin{equation}        \bar{\Omega}=\mathbb{E}_{\bar{a},\bar{b},\bar{c}}=\Omega \cup\Omega_{\mathrm{sea}},
\end{equation}
with $\bar{a} = s_L a$, $\bar{b} = s_L b$, and $\bar{c} = s_L c$.

Within $\bar{\Omega}$, the original boundary $\partial \Omega$ becomes an interior interface, while the outermost sea layer forms the new boundary $\partial \bar{\Omega}$. The sea serves as a volumetric buffer region for density diffusion and improves the stability of the free-boundary optimization process.

The algorithm for the construction of the sea is summarized in Algorithm~\ref{alg:construction_sea}.
\begin{algorithm}[h]
\caption{Construction of the volumetric sea}
\label{alg:construction_sea}

\KwIn{A tetrahedral mesh of the ellipsoidal domain $\Omega$, the number of layers $L$, and the layer ratio $\kappa$.}
\KwOut{An extended tetrahedral mesh $\bar{\Omega}$ containing a volumetric sea.}

Extract the boundary triangulation $\mathcal{F}_{\partial\Omega}$\;

Compute the mean boundary edge length $\bar{l}$ and set $h=\kappa\bar{l}$\;

Generate $L$ concentric ellipsoidal layers by radial scaling\;

Connect every pair of adjacent layers to form triangular prisms\;

Subdivide each triangular prism into tetrahedra using a standard prism decomposition\;

Assemble the sea mesh and merge it with the original tetrahedral mesh\;

The resulting sea-embedded domain is given by $\bar{\Omega}$\;

\end{algorithm}

\subsubsection{Free-boundary update}\label{sec:free_bdy_update}
Next, we update the parametrization on this sea-embedded domain $\bar \Omega = \Omega^n \cup \Omega^n_{\mathrm{sea}}$. {At initialization, the density in the auxiliary sea $\Omega_{\mathrm{sea}}^0$ is set to the mean value of the initial density on the original domain.}

The update on the original domain follows the same coupled strategy as in the prescribed ellipsoidal model. We compute the local shape and density descents and combine them to deform the mapping. Note that we omit the projection step for the vertices on the original boundary. Therefore, the original domain is treated as an interior part and can move freely in $\bar \Omega^n$. For the auxiliary sea domain, we focus solely on the density-driven deformation and omit the local shape distortion. Moreover, we apply a projection step for the vertices on the outside boundary of the sea-embedded domain to project them back to the fixed outer ellipsoid $\partial \bar \Omega$.

\begin{remark}
    The auxiliary sea region is not included in the local shape distortion term. {Its initial density is the mean value of the initial density on the original domain. During the iterations, the diffusion step updates the density on the entire extended domain $\bar \Omega$, including the auxiliary sea. The updated sea density is retained for the next iteration and is not reset to the mean density of the current original domain. Thus, the sea density may develop spatial variations during the evolution.}
\end{remark}

\subsubsection{Summary}
In summary, the sea-embedded free-boundary model updates the original target domain together with an auxiliary sea region. The original tetrahedra follow the coupled shape-density update, whereas the sea tetrahedra are used only for the density-driven update. The fixed outer ellipsoid prevents uncontrolled expansion, while the sea region provides auxiliary degrees of freedom without introducing an additional energy contribution. The complete sea-embedded free-boundary method is summarized in Algorithm~\ref{alg:free_bdy}.

\begin{algorithm}[h]
\caption{Sea-embedded free-boundary volumetric parameterization}
\label{alg:free_bdy}

\KwIn{A simply connected 3-manifold $\mathcal{M}$, a prescribed mass distribution, weight parameters $\alpha$ and $\beta$, the initial ellipsoidal radii $(a_0,b_0,c_0)$, the sea parameters $L$ and $\kappa$, the time-step size $\delta t$, the stopping parameter $\epsilon$, and the maximum number of iterations allowed $n_{\max}$.}

\KwOut{A free-boundary volumetric parameterization
$f:\mathcal{M}\to\mathbb{R}^3$.}

Compute an initial ellipsoidal map
$f^0:\mathcal{M}\to\Omega^0=\mathbb{E}_{a_0,b_0,c_0}$\;

Apply the construction scheme in
Section~\ref{sec:create_sea} to obtain the sea-embedded domain $
\bar{\Omega}
=
\Omega^0\cup\Omega_{\mathrm{sea}}^0$\;

Initialize the density on $\bar{\Omega}$\;

Set $n=0$\;

\Repeat{$\lVert f^n-f^{n-1}\rVert<\epsilon$
\ or \ $n\geq n_{\max}$}{

Compute the diffusion-driven update on $\bar{\Omega}$\;

Compute the local shape-driven update on the original domain
$\Omega^n$\;

Combine the diffusion-driven and shape-driven directions to obtain the updated map on $\bar{\Omega}$\;

Apply the geometric correction scheme to $\bar{\Omega}$\;

Set $n=n+1$\;
}

Remove the auxiliary sea\;

The resulting free-boundary volumetric parameterization is given by $f=f^n$\;

\end{algorithm}

\section{Experimental results}\label{sec:experimental_results}
In this section, we present experimental results to demonstrate the effectiveness of the proposed volumetric parameterization framework. The proposed algorithms are implemented in MATLAB R2021a. All experiments are conducted on a Windows computer with an Intel(R) Core(TM) i9-12900 2.40 GHz processor and 32 GB of memory. {The parameters used in the proposed algorithms are set as follows: $\alpha=0.1$, $\beta=1$, $C=10$, and $K_T=10$. These parameters control the relative weights of the shape and mass updates, the shape update step, and the geometric correction threshold, respectively.} All 3-manifolds are discretized in the form of tetrahedral meshes.

\subsection{Prescribed ellipsoidal model}
\begin{figure}[t!]
    \centering
    \includegraphics[width=1\textwidth]{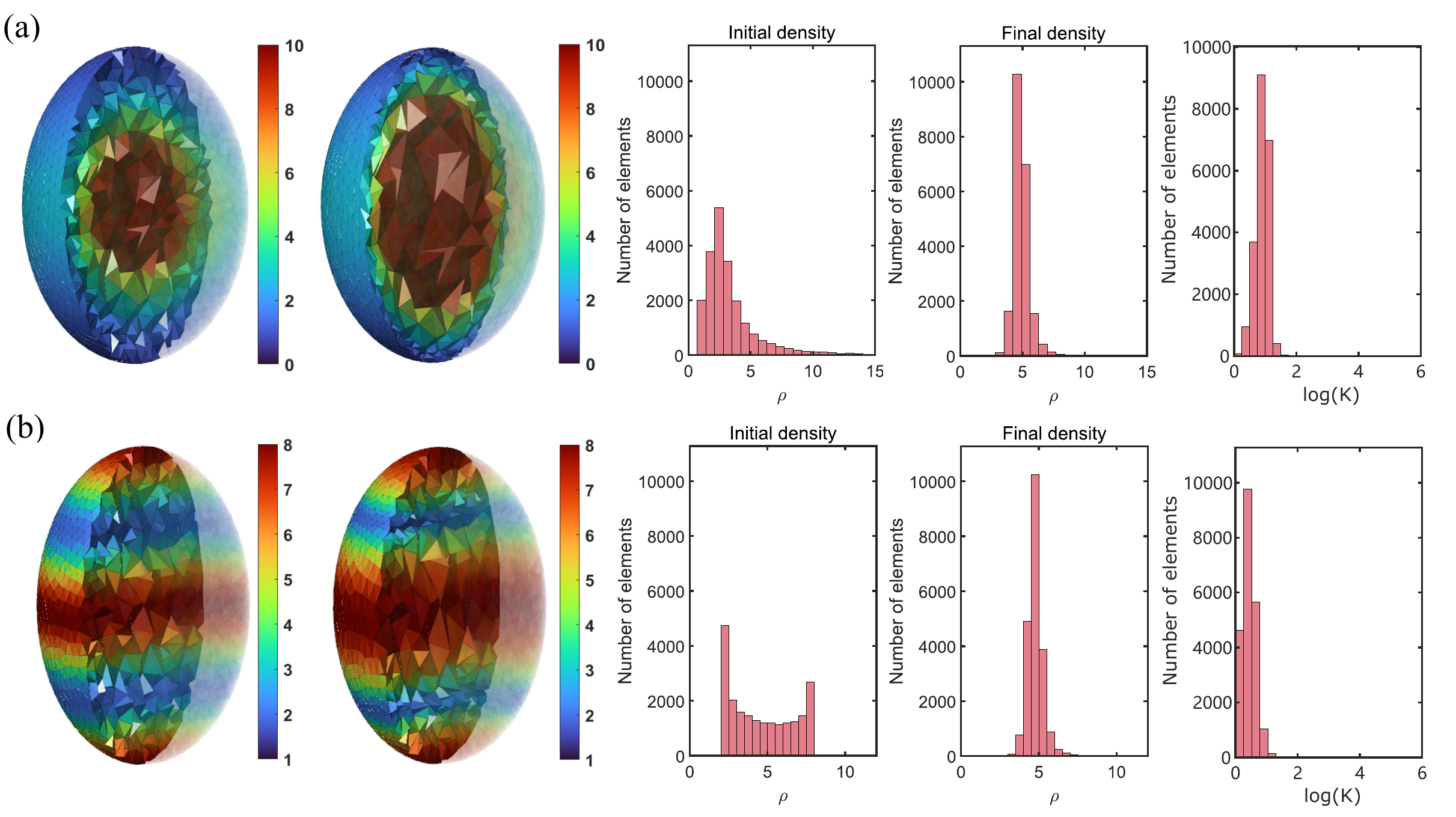}
    \caption{\textbf{Ellipsoidal mass-preserving quasi-conformal maps of solid ellipsoids.} Each row shows one example. (a) An example with a continuous input density that varies with the distance to the origin. (b) An example with a continuous input density that varies with the $z$-coordinate. Left to right: The initial solid ellipsoid color-coded with the initial density, the final result color-coded with the initial density, the histogram of the initial density, the histogram of the final density, and the histogram of $\log(K)$. }
    \label{fig:ellipsoid_fix_shape}
\end{figure}

We first test our proposed prescribed ellipsoidal method by mapping some solid ellipsoids with different density distributions. In the examples shown in Fig.~\ref{fig:ellipsoid_fix_shape}, we define different mass on the solid ellipsoid to obtain different initial densities. { For each tetrahedron $T$, let $v_T^c=(x_T,y_T,z_T)$ denote its barycenter in the initial solid ellipsoid, and let $r_T=\|v_T^c\|_2$. We prescribe the mass $m_T$ of each
tetrahedron according to the following formulas. The prescribed mass distributions in Fig.~\ref{fig:ellipsoid_fix_shape}(a) and (b) are defined as $m_T^{(a)} = \exp(-3r_T)$ and $m_T^{(b)} = \bigl(3\cos(4z_T)+5\bigr)$.} It can be observed that the initial densities are highly non-uniform in both examples. We then apply the proposed method and obtain the prescribed ellipsoidal mapping results as shown in the figure. Compared with the initial maps, the results enlarge the high-density regions while significantly shrinking the low-density regions. From the density histograms, we can see that the density is effectively equalized. Moreover, the histograms of $\log(K)$ show that the results effectively control local shape.
 
\begin{figure}[t!]
    \centering
    \includegraphics[width=1\textwidth]{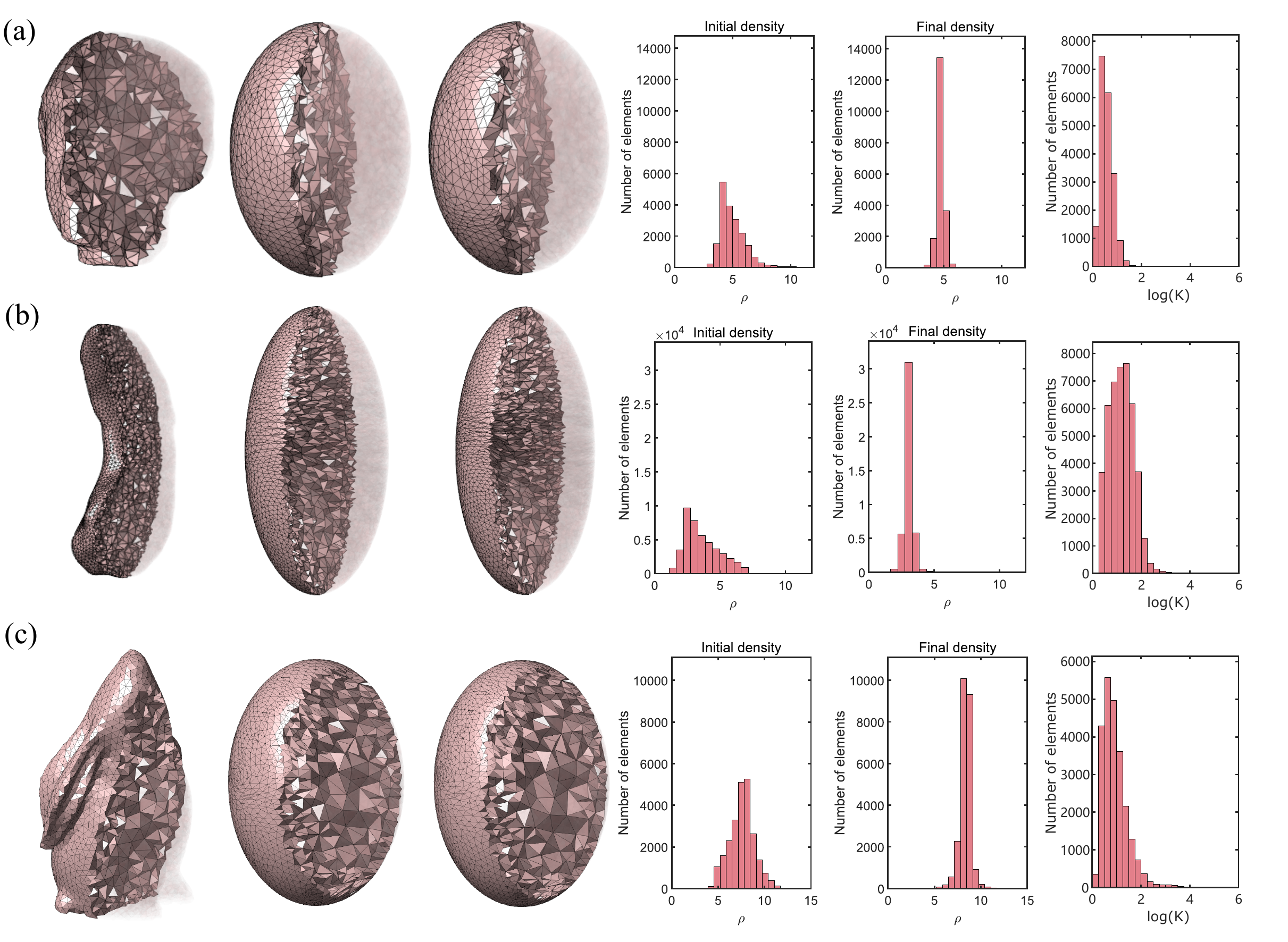}
    \caption{\textbf{Ellipsoidal volumetric parameterization of simply connected 3-manifolds obtained by our method.} Each row shows one example. (a)~The Igea model. (b)~The Hippocampus model. (c)~The Rabbit model. Left to right: The original 3-manifold, the initial solid ellipsoid, the final result, the histogram of the initial density, the histogram of the final density, and the histogram of $\log(K)$.}
    \label{fig:real_fix_shape}
\end{figure}

Then we consider the prescribed ellipsoidal mass-preserving quasi-conformal parameterizations for general simply connected manifolds. In Fig.~\ref{fig:real_fix_shape}, we prescribe different mass distributions on
the input tetrahedral meshes. The prescribed masses for the Igea, Hippocampus, and Rabbit models are given by
\begin{equation}
    m_T^{\mathrm{Igea}} = \text{Vol}(T_{\text{Igea}}),\ m_T^{\mathrm{Hippocampus}} = e^{r_T},\ m_T^{\mathrm{Rabbit}} = 3(4-(r_T+0.2)^2).
\end{equation}
Here, $\text{Vol}(T_{\text{Igea}})$ is the volume of tetrahedrons in the original Igea model. The initial density histograms in  Fig.~\ref{fig:real_fix_shape} show that the density distortions are large in the initial ellipsoidal parameterization for all examples. Then we apply the proposed method and obtain the mapping results. The final density histograms in Fig.~\ref{fig:real_fix_shape} are highly concentrated, indicating that the prescribed mass is more uniformly distributed over the target ellipsoids. Furthermore, the local shape distortion histograms in Fig.~\ref{fig:real_fix_shape} concentrate near zero, which shows that the local shape distortions are controlled effectively. 

For a more quantitative analysis of our prescribed ellipsoidal volumetric parameterization method, we record the radii of the target ellipsoidal domain, variance of the initial density, variance of the final density, mean and standard deviations of the 3D quasi-conformal dilation $K$, and number of overlaps for mapping different models (see Table~\ref{tab:model1_fix_shape_results}). Compared with the initial density, the variance of the final density is significantly reduced, which indicates that our proposed method effectively equalizes the density. Besides, the mean and standard deviation of $K$ further characterize the average level and variability of local shape distortion, respectively. Moreover, from the number of overlaps, we can see that the mappings are all bijective.

\textbf{\begin{table}[t!]
\centering
\caption{\textbf{The performance of our prescribed ellipsoidal model.} For each example, we record the number of tetrahedral elements, the radii $(a,b,c)$, the variance of the normalized initial density $\widetilde{\rho}_{0} = \frac{\rho_0}{\text{Mean}(\rho_0)}$, the normalized final density $\widetilde{\rho}_{1} = \frac{\rho_1}{\text{Mean}(\rho_1)}$, the mean and standard deviations of the 3D quasi-conformal dilation $K$.
}
\label{tab:model1_fix_shape_results}
\resizebox{1\linewidth}{!}{$
\begin{tabular}{lccccccc}
\toprule
Example 
& \# Elements
& Radii $(a,b,c)$
& $\operatorname{Var}(\widetilde{\rho}_{0})$
& $\operatorname{Var}(\widetilde{\rho}_{1})$
& Mean$(K)$
& SD$(K)$
& \# Overlaps \\
\midrule
Ellipsoid 1 & 21294 & $(1,1,1.4)$ & 0.7125 & 0.0220 & 2.5205 & 0.5254 & 0 \\
Ellipsoid 2 & 21294 & $(1,1,1.5)$ & 0.1968 & 0.0174 & 1.5927 & 0.3758 & 0 \\
Igea        & 19539 & $(1,1,1.4)$ & 2.6446 & 0.1005 & 1.8369 & 1.0277 & 0 \\
Hippocampus & 43833 & $(1,1,2)$   & 0.1315 & 0.0440 & 3.8117 & 5.8654 & 0 \\
Rabbit      & 23954 & $(1,1,1.4)$ & 0.0303 & 0.0176 & 3.2448 & 7.3817 & 0 \\
\bottomrule
\end{tabular}
$}
\end{table}}

\begin{table}[t!]
\centering
\caption{\textbf{The results of the Igea model obtained by our prescribed ellipsoidal algorithm with different elliptic radii.} Here, for different choices of elliptic radii $(a,b,c)$ of the target solid ellipsoidal domain, we record the mean and standard deviations of volume distortion $|\mathcal{D}_{\mathrm{vol}}|$, the mean and standard deviations of 3D quasi-conformal dilation $K$, and the number of overlaps.}
\label{tab:igea_radii_comparison}
\resizebox{0.85\linewidth}{!}{$
\begin{tabular}{cccccccc}
\toprule
$a$ & $b$ & $c$ 
& Mean$(|\mathcal{D}_{\mathrm{vol}}|)$ 
& SD$(|\mathcal{D}_{\mathrm{vol}}|)$ 
& Mean$(K)$
& SD$(K)$
& \# Overlaps \\
\midrule
1 & 1.0 & 1.0 & 0.0492 & 0.0768 & 1.9176 & 0.8725 & 0 \\
  & 0.8 & 1.0 & 0.0500 & 0.0806 & 2.1165 & 1.0539 & 0 \\
  & 1.2 & 1.0 & 0.0504 & 0.0754 & 1.8795 & 0.8053 & 0 \\
  & 1.4 & 1.0 & 0.0513 & 0.0751 & 1.9021 & 0.7870 & 0 \\
  & 1.0 & 0.8 & 0.0505 & 0.0783 & 2.1334 & 0.9019 & 0 \\
  & 1.0 & 1.2 & 0.0501 & 0.0775 & 1.8314 & 0.9255 & 0 \\
  & 1.0 & 1.4 & 0.0508 & 0.0781 & 1.8369 & 1.0277 & 0 \\
  & 1.0 & 1.6 & 0.0526 & 0.0862 & 1.9461 & 3.1371 & 0 \\   
  & 0.6 & 1.2 & 0.0574 & 0.0893 & 2.6091 & 1.5506 & 0 \\
  & 0.8 & 1.2 & 0.0516 & 0.0816 & 2.0803 & 1.1362 & 0 \\
  & 1.2 & 1.6 & 0.0516 & 0.0833 & 1.7724 & 1.8767 & 0 \\
  & 1.4 & 1.8 & 0.0535 & 0.0847 & 1.8009 & 1.1723 & 0 \\
\bottomrule
\end{tabular}
$}
\end{table}

It is natural to ask how the choices of the radii $(a,b,c)$ in the energy \eqref{eq:prescribed_ellispoid} will affect the mapping results in the proposed prescribed ellipsoidal method. Here, we consider the Igea model for volume-preserving quasi-conformal parameterizations. Specifically, we use different values of $(a,b,c)$ and analyze the mapping results in terms of volume distortions, local shape error, and bijectivity. In Table~\ref{tab:igea_radii_comparison}, we report the mean and standard deviation of volume distortion $\mathcal{D}_{\mathrm{vol}}$, the mean and standard deviation of 3D quasi-conformal dilation $K$, as well as the number of overlaps. The volume distortion $\mathcal{D}_{\mathrm{vol}}$ for each tetrahedral element $T$ is given by
\begin{equation}
    \mathcal{D}_{\mathrm{vol}}(T) = \log\left( \frac{\mathrm{Volume}(f(T))/\sum_{T^{'}\in \mathcal{T}}\mathrm{Volume}(f(T^{'}))}{\mathrm{Volume}(T)/\sum_{T^{'}\in \mathcal{T}}\mathrm{Volume}(T^{'})} \right).
\end{equation}
Here, $f$ denotes the solid ellipsoidal parameterization and $\mathcal{T}$ is the set of tetrahedral elements. The above normalization factors and denominator ensure that a perfectly volume-preserving result would yield $\mathcal{D}_{\mathrm{vol}} = 0$. From Table~\ref{tab:igea_radii_comparison}, we can observe that the mapping results can reduce the volume distortions effectively for all combinations of the elliptic radii. Moreover, the local shape distortion is more sensitive to the elliptic radii. With different radii, the mean value of $K$ varies from $1.7724$ to $2.6091$, and the standard deviation varies from $0.7870$ to $3.1371$. Thus, although the prescribed model effectively redistributes the mass for different ellipsoidal targets, the choice of target radii may have a notable influence on the geometric quality of the map. The number of overlaps is zero for all test examples, demonstrating the effectiveness of our geometric correction scheme.

\subsection{Volume-normalized adaptive ellipsoidal model}
\begin{figure}[t!]
    \centering
    \includegraphics[width=1\textwidth]{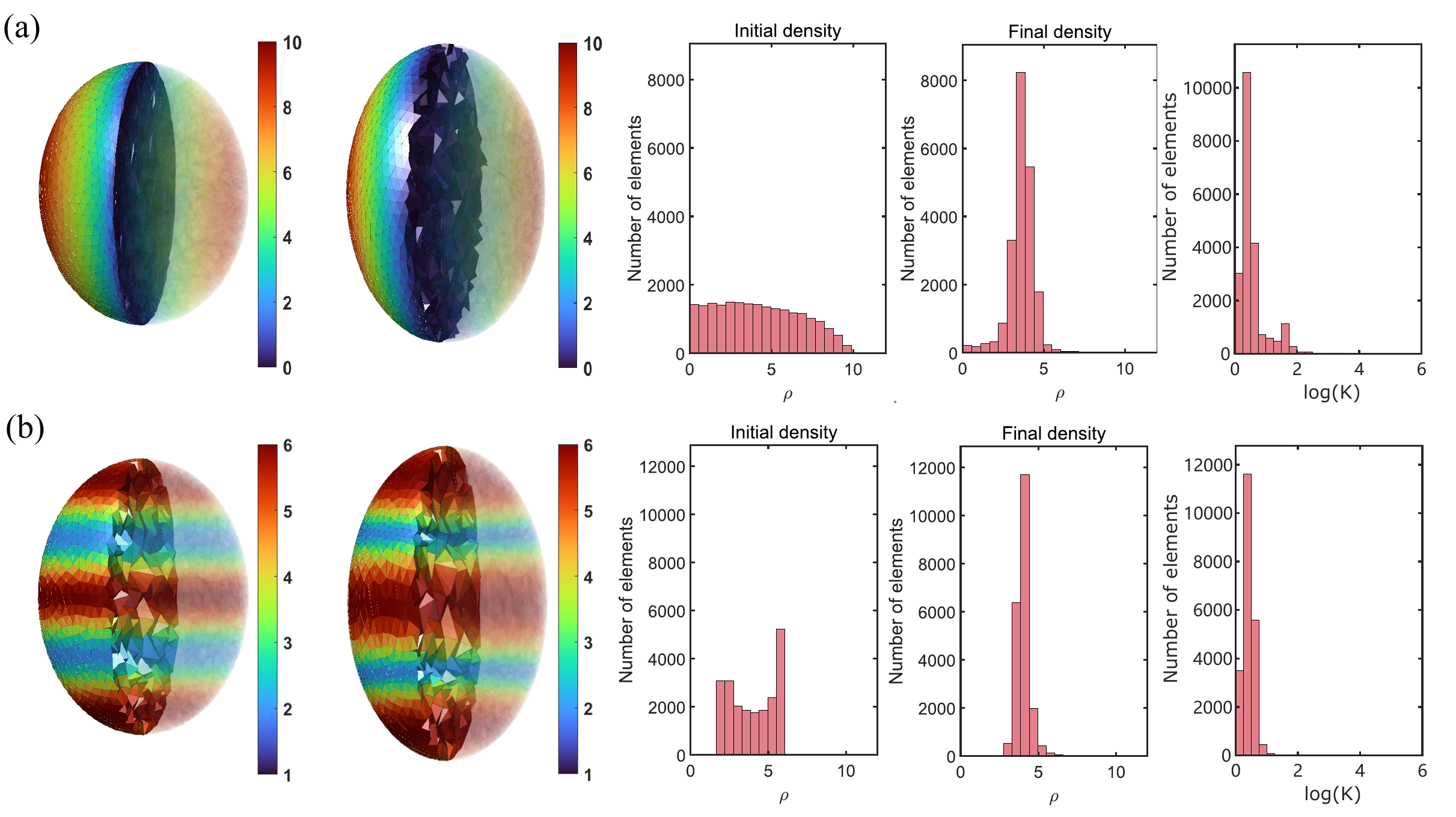}
    \caption{\textbf{Volume-normalized adaptive ellipsoidal mass-preserving quasi-conformal maps of solid ellipsoids.} Each row shows one example. (a)~An example with a continuous input density varying with the $y$-coordinate. (b)~An example with a continuous input density varying with the $z$-coordinate. Left to right: The initial solid ellipsoid color-coded with the initial density, the final result color-coded with the initial density, the histogram of the initial density, the histogram of the final density, and the histogram of $\log(K)$.}
    \label{fig:ellipsoid_optimal_shape}
\end{figure}

We next evaluate the volume-normalized adaptive ellipsoidal model. Unlike the above method, this model updates both the volumetric mapping and the ellipsoidal radii. As shown in Fig.~\ref{fig:ellipsoid_optimal_shape}(a), the initial mass distributions are defined based on the $y$-coordinates, with the input density varying from $0.5$ to $10$. 
{ The corresponding mass assignment is $m_T^{(a)}=10|y_T+0.1|.$} After applying our method, the final ellipsoid exhibits a more slender shape. Furthermore, the density histograms indicate that our method reduces the density distortions efficiently. Meanwhile, the histogram of $\log(K)$ shows that the resulting mapping effectively controls local shape distortion. Fig.~\ref{fig:ellipsoid_optimal_shape}(b) shows another example with the initial density varying with the $z$-coordinate. 
{ The mass assigned to each tetrahedron is $ m_T^{(b)} = 2\cos(2\pi z_T)+4$.} The resulting ellipsoid is stretched along the $z$-direction, while the high-density domains expand and the low-density domains shrink. The density and local shape histograms show that the resulting mapping equalizes the density while maintaining low shape distortions.

\begin{figure}[t!]
    \centering
    \includegraphics[width=1\textwidth]{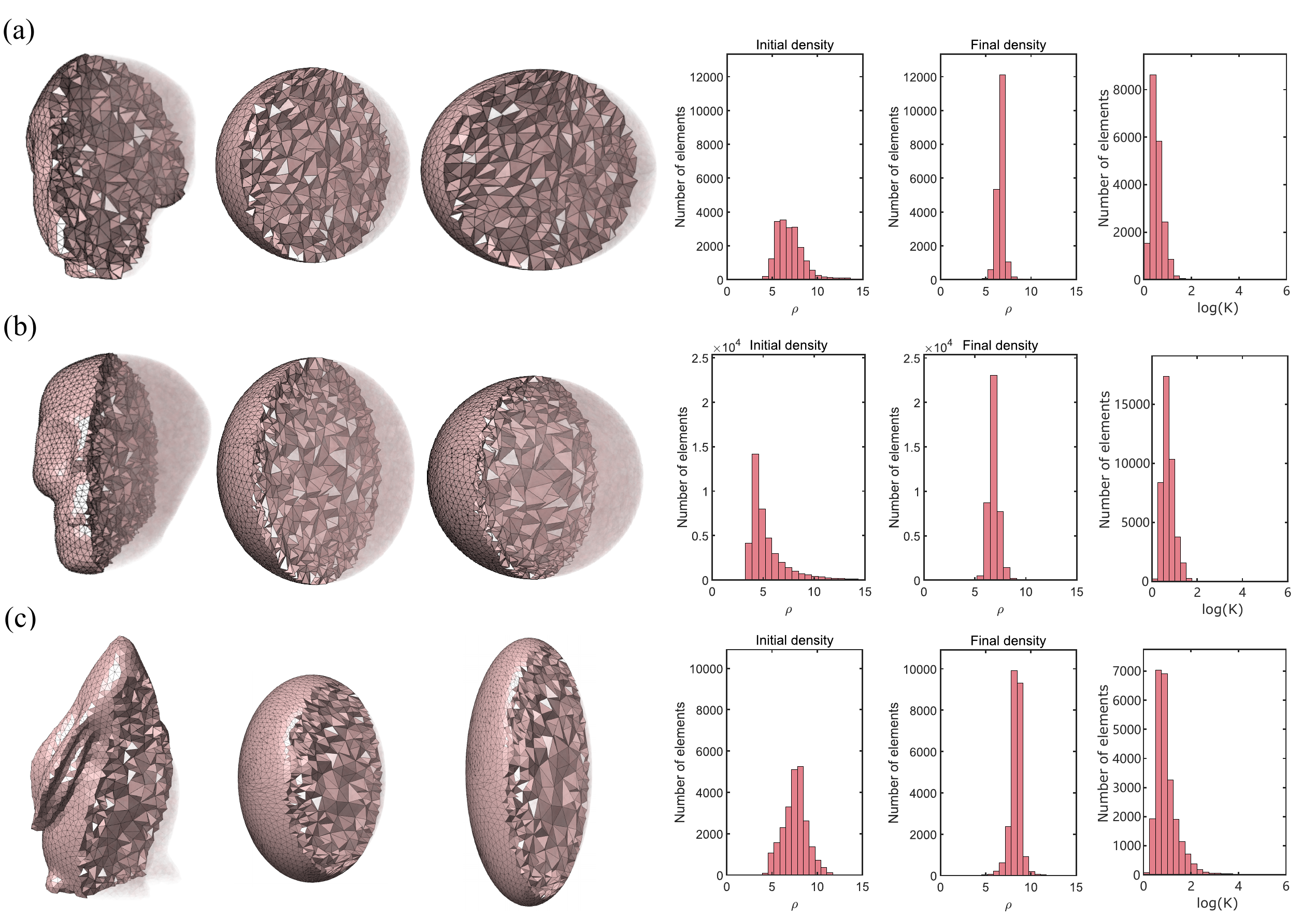}
    \caption{\textbf{Volume-normalized adaptive ellipsoidal volumetric parameterization of simply connected 3-manifolds obtained by our method.} Each row shows one example. (a) The Igea model. (b)~The Skull model. (c)~The Rabbit model. Left to right: The original 3-manifold, the initial solid ellipsoid, the final result, the histogram of the initial density, the histogram of the final density, and the histogram of $\log(K)$.}
    \label{fig:real_optimal_shape}
\end{figure}

Next, we consider computing adaptive ellipsoidal volumetric parameterizations for simply connected 3-manifolds. For each example in Fig.~\ref{fig:real_optimal_shape}, we first map the 3-manifold to a solid ellipsoid with a specific mass distribution and arbitrary radii. 
The prescribed masses for the Igea, Skull, and Rabbit models are given by
\begin{equation}
    m_T^{\mathrm{Igea}} = \text{Vol}(T_{\text{Igea}}),\ m_T^{\mathrm{Skull}} = 20e^{-2r_T},\ m_T^{\mathrm{Rabbit}} = 3(4-(r_T+0.2)^2).
\end{equation}
Then, we apply our volume-normalized adaptive ellipsoidal algorithm to obtain the ellipsoidal volumetric parameterizations. Compared with the initial mappings, we can see that the shapes of the final results are changed significantly. Moreover, the density histograms show that the resulting parameterizations are highly mass-preserving, while the histograms of $\log(K)$ demonstrate that local shape distortions are effectively controlled.

Table~\ref{tab:model2_optimal_shape_results} provides a more quantitative analysis of our volume-normalized adaptive ellipsoidal method. We record the initial and final radii, variance of the initial and final density, mean and standard deviation of 3D quasi-conformal dilation, and the number of overlaps. The differences between the initial and final radii indicate that our method effectively adjusts the ellipsoidal shape based on the energy formula. {The mean and standard deviation of $K$ characterize its average level and variability, respectively. To further examine the distribution, we report the median, interquartile range (IQR), and $C_{1\%}=\frac{\sum_{i\in I_K}(K_i-\text{Mean}(K))^2}{\sum_{i=1}^{N}(K_i-\text{Mean}(K))^2}$, which measures the contribution of the $1\%$ of elements with the largest $K$ values (indexed by $I_K$) to the total squared deviation of $K$ over all $N$ tetrahedral elements. For Ellipsoid 3 and Rabbit, $C_{1\%}$ reaches 0.9758 and 0.9445, respectively, indicating that their large standard deviations are primarily driven by a small fraction of highly distorted elements.} Finally, the number of overlaps is $0$ for all examples, indicating that all resulting mappings are folding-free.

\begin{table}[t!]
\centering
\caption{\textbf{The performance of the volume-normalized adaptive ellipsoidal model.} For each example, we record the number of tetrahedral elements, the initial radii $(a_0,b_0,c_0)$, the final radii $(a_1,b_1,c_1)$, the variance of the normalized initial density $\widetilde{\rho}_{0} = \frac{\rho_0}{\text{Mean}(\rho_0)}$, the normalized final density $\widetilde{\rho}_{1} = \frac{\rho_1}{\text{Mean}(\rho_1)}$, the mean, standard deviation, median, {the interquartile range (IQR) of the 3D quasi-conformal dilation $K$, $C_{1\%}$,}  and the number of overlaps.
}
\label{tab:model2_optimal_shape_results}
\resizebox{\linewidth}{!}{$
\begin{tabular}{lccccccccccc}
\toprule
Example 
& \# Elements
& Initial radii $(a_0,b_0,c_0)$ 
& Final radii $(a_1,b_1,c_1)$
& $\operatorname{Var}(\widetilde{\rho}_{0})$
& $\operatorname{Var}(\widetilde{\rho}_{1})$
& Mean$(K)$
& SD$(K)$
& {Median$(K)$}
& {IQR$(K)$}
& {$C_{1\%}$}
& \# Overlaps \\
\midrule
Ellipsoid 3 
& 21294
& $(1,1,1.5)$ 
& $(1.1395,0.9319,1.4126)$
& 0.3684 & 0.0915 & 2.3578 & 8.3280 & {1.5429} & {0.3889} & {0.9758} & 0 \\

Ellipsoid 4 
& 21294
& $(1,0.6,1.2)$ 
& $(0.9418,0.5651,1.3530)$
& 0.1272 & 0.0162 & 1.5275 & 0.3219 & {1.4612} & {0.3543} & {0.4302} & 0 \\

Igea 
& 19539
& $(1,1,1)$ 
& $(0.8891,1.1248,1.0000)$
& 2.6008 & 0.0637 & 1.7976 & 0.8411 & { 1.6264} & { 0.5236} & {0.7219} & 0 \\

Skull 
& 41938
& $(1,1,1.1)$ 
& $(0.9412,1.1008,1.0611)$
& 0.1638 & 0.0066 & 2.1235 & 0.7820 & {1.9742} & {0.6269} & {0.4786} & 0 \\

Rabbit
& 23954
& $(1,1,1.4)$ 
& $(0.8708,0.8708,1.8462)$
& 0.0303 & 0.0080 & 3.2706 & 6.4489 & {2.3366} & {1.0018} & {0.9445} & 0 \\
\bottomrule
\end{tabular}
$}
\end{table}

\begin{figure}[t!]
    \centering
    \includegraphics[width=1\textwidth]{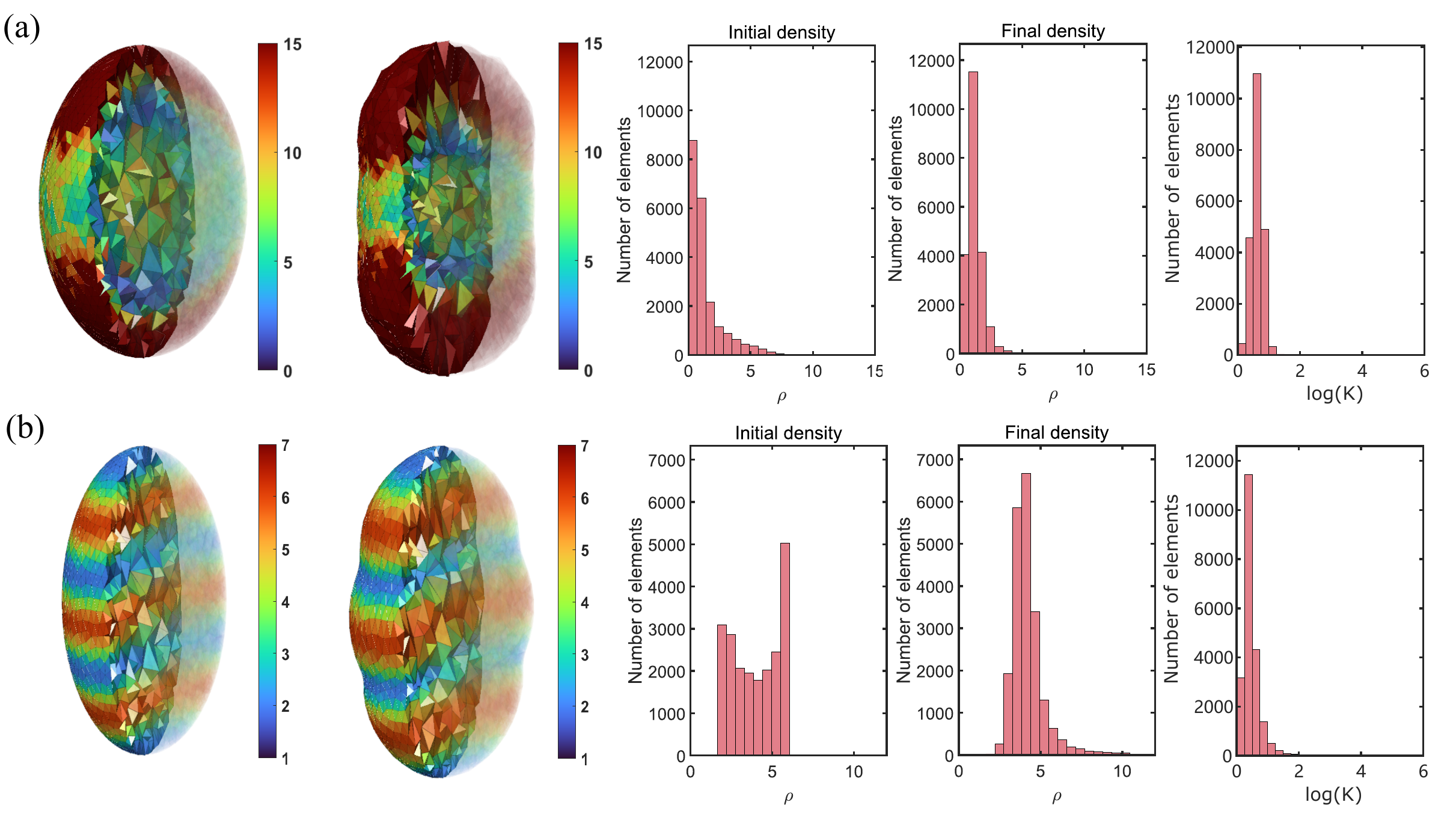}
    \caption{\textbf{Free-boundary density-equalizing quasi-conformal maps of ellipsoidal 3-manifolds.} Each row shows one example. (a) An example with a continuous input density varying with the distance to the origin. (b) An example with a continuous input density depending on a linear combination of the $y$- and $z$-coordinates. Left to right: The initial solid ellipsoid color-coded with the initial density, the final result color-coded with the initial density, the histogram of the initial density, the histogram of the final density, and the histogram of $\log(K)$.}
    \label{fig:ellipsoid_free_bdy}
\end{figure} 

\subsection{Sea-embedded free-boundary model}
Then, we test our sea-embedded free-boundary model for computing free-boundary mass-preserving quasi-conformal mappings. {Here, the number of auxiliary sea layers and the layer thickness parameter are set to $L=3$ and $\kappa=2$, respectively.} In Fig.~\ref{fig:ellipsoid_free_bdy}, we define different mass distributions on the solid ellipsoid and prescribe different continuous input density distributions. { The prescribed masses in Fig.~\ref{fig:ellipsoid_free_bdy}(a) and (b) are given by $m_T^{(a)} = (4\sin(5r_T)+5)$ and $m_T^{(b)} = 2\cos\bigl(2\pi(0.5y_T+z_T)\bigr)+4$.} After applying our free-boundary method to these ellipsoidal examples, it can be observed that the geometry of the resulting mappings changes significantly, and the final domains are no longer ellipsoidal. Moreover, high-density regions are enlarged while low-density regions are shrunk, resulting in equalized final density distributions. Concurrently, the histograms of $\log(K)$ demonstrate that local shape distortions remain well-controlled.

\begin{figure}[t!]
    \centering
    \includegraphics[width=1\textwidth]{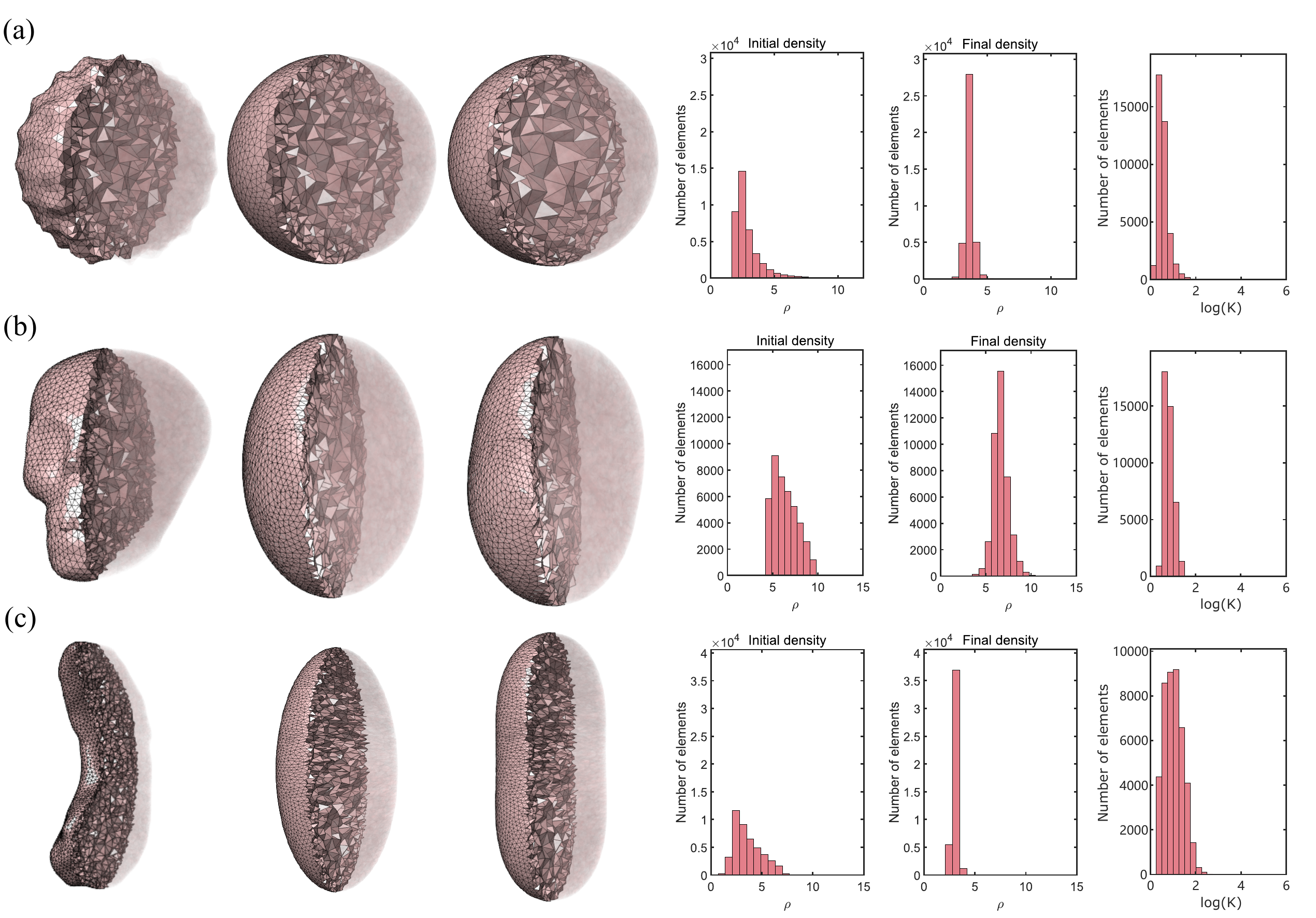}
    \caption{\textbf{Free-boundary volumetric parameterization of simply connected 3-manifolds obtained by our method.} Each row shows one example. (a) The Multi-lobed ball model. (b) The Skull model. (c) The Hippocampus model. Left to right: The original 3-manifolds, the initial solid ellipsoid, the final result, the histogram of the initial density, the histogram of the final density, and the histogram of $\log(K)$.}
    \label{fig:real_free_bdy}
\end{figure}

In Fig.~\ref{fig:real_free_bdy}, we present the free-boundary volumetric parameterizations for some simply connected 3-manifolds. 
The prescribed masses for the Multi-lobed Ball, Skull, and Hippocampus models are given by
\begin{equation}
    m_T^{\mathrm{Multi\text{-}lobed}} = 15e^{-2r_T},\ m_T^{\mathrm{Skull}} = 20e^{-2r_T},\ m_T^{\mathrm{Hippocampus}} = e^{r_T}.
\end{equation} 
In all examples, the proposed method yields mass-preserving quasi-conformal mappings. Specifically, the shape of the target domain is optimized effectively during the iteration process under the free-boundary condition. Compared with the initial density, the final density becomes highly equalized. Moreover, the local shape distortions are optimized due to the shape change. Table~\ref{tab:model3_free_boundary_results} summarizes the initial radii, variance of normalized initial and final densities, mean and standard deviation of $K$, and the number of overlaps. We can see that the proposed free-boundary method can reduce density distortion significantly while maintaining low local shape distortion. This demonstrates that our algorithm is capable of computing free-boundary mass-preserving quasi-conformal parameterizations for a wide range of simply connected 3-manifolds.

\begin{table}[t!]
\centering
\caption{\textbf{The performance of the sea-embedded free-boundary model.}
For each example, we record the number of tetrahedral elements, the initial radii $(a,b,c)$, the variance of the normalized initial density $\widetilde{\rho}_{0} = \frac{\rho_0}{\text{Mean}(\rho_0)}$, the normalized final density $\widetilde{\rho}_{1} = \frac{\rho_1}{\text{Mean}(\rho_1)}$, the mean and standard deviations of the 3D quasi-conformal dilation $K$, and the number of overlaps.
}
\label{tab:model3_free_boundary_results}
\resizebox{1\linewidth}{!}{$
\begin{tabular}{lccccccc}
\toprule
Example 
& \# Elements
& Initial $(a,b,c)$
& $\operatorname{Var}(\widetilde{\rho}_{0})$
& $\operatorname{Var}(\widetilde{\rho}_{1})$
& Mean$(K)$
& SD$(K)$
& \# Overlaps \\
\midrule
Ellipsoid 5 
& 21294
& $(1,1,1.5)$ 
& 1.0162 & 0.3161 & 1.9055 & 0.3565 & 0 \\

Ellipsoid 6 
& 21294
& $(1,0.6,1.6)$ 
& 0.1242 & 0.0873 & 1.6629 & 0.7046 & 0 \\

Multi-lobed ball
& 38917
& $(1,1,1)$ 
& 0.1438 & 0.0091 & 1.8377 & 2.8455 & 0 \\

Skull 
& 41938
& $(1,1,1.6)$ 
& 0.0422 & 0.0020 & 2.2285 & 0.6750 & 0 \\

Hippocampus
& 43833
& $(1,1,2)$ 
& 0.1315 & 0.0173 & 3.0606 & 2.8963 & 0 \\
\bottomrule
\end{tabular}
$}
\end{table}

\subsection{Comparison of the proposed models}\label{sec:comparisom_result}
\begin{figure}[t!]
    \centering
    \includegraphics[width=1\textwidth]{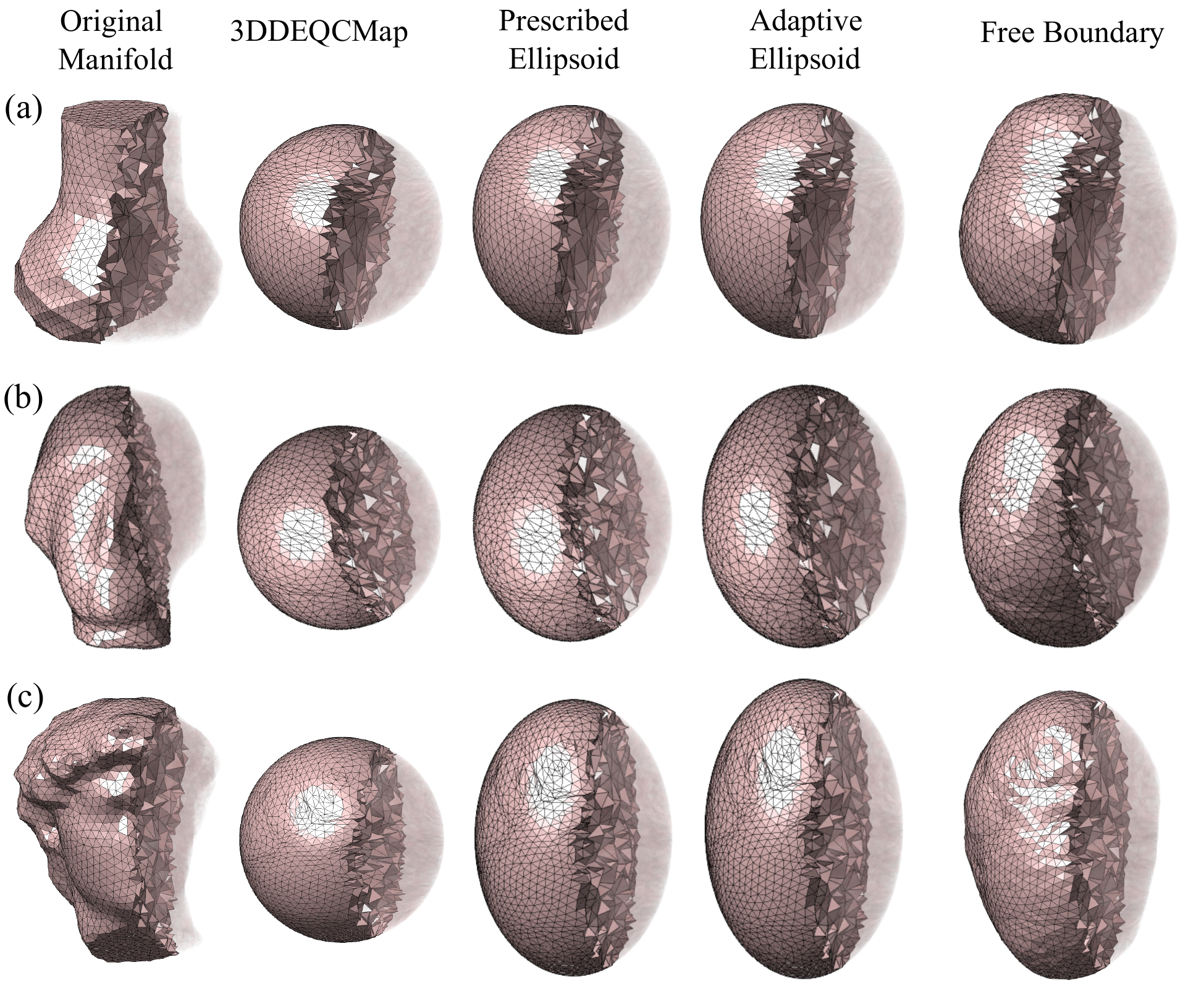}
    \caption{{\textbf{Comparison of volumetric parameterizations under different target-domain settings.} (a) The Vase model. (b) The Igea model. (c) The David model. From left to right: the original 3-manifold, the unit-ball parameterization obtained by 3DDEQCMap~\cite{lyu2025volumetric}, and the results of the prescribed ellipsoidal, volume-normalized adaptive ellipsoidal, and sea-embedded free-boundary methods.}}
    \label{fig:model_comparison}
\end{figure}

{The previous subsections have examined the three proposed models separately. We now compare their results under different target-domain settings. Fig.~\ref{fig:model_comparison} provides the comparison of different volumetric parameterization methods for the Vase, Igea, and David models. For each tetrahedron $T$ in the original mesh, we assign the prescribed mass $m_T=\operatorname{Vol}(T)$. Assigning each tetrahedron its original volume as mass makes preservation of relative tetrahedral volumes the density-equalization objective. For each example, we show the original 3-manifold, the unit-ball parameterization obtained by 3DDEQCMap~\cite{lyu2025volumetric}, and the results of our three proposed models. The prescribed ellipsoidal model maps the input manifold onto a fixed ellipsoid, while the volume-normalized adaptive ellipsoidal model optimizes the ellipsoidal radii. The sea-embedded free-boundary model further relaxes the ellipsoidal constraint, allowing more flexible target geometries. These results illustrate the different target-domain shapes and volumetric mesh structures obtained by different methods.}

For a more quantitative analysis of our parameterization framework, in Table~\ref{tab:three_model_comparison} we record the result radii, the variance of the final density, the mean and standard deviation of $K$, and the number of overlaps for different target domain settings. For comparison, we also include the solid ball parameterization results~\cite{lyu2025volumetric} as a baseline. Note that for the case of a prescribed ellipsoid, the parameterization performance depends on the choice of the radii and hence is notably different from the solid ball case. Moreover, compared with the solid ball results, the adaptive ellipsoidal and free-boundary models reduce both density and local shape distortions across all examples. Among the three proposed models, the results demonstrate that relaxing the target-domain constraint generally provides additional flexibility for balancing mass and local shape distortions. In particular, the adaptive ellipsoidal and free-boundary methods consistently reduce the final density variance compared with the prescribed ellipsoidal method. The free-boundary method also substantially reduces the standard deviation of the quasi-conformal dilation $K$ in all three examples, which indicates that this method controls local shape distortions effectively. Moreover, the number of overlaps is zero, indicating that all result mappings are folding-free. This comparison highlights the flexibility of our proposed framework for handling 3-manifolds with different geometries. { To further examine the remaining density variation, we also report the median, interquartile range (IQR), and $C_{1\%}^{\rho}$ of the normalized final density in Table~\ref{tab:three_model_comparison}. $C_{1\%}^{\rho}=\frac{\sum_{i\in I_\rho}(\widetilde{\rho}_{1,i}-1)^2}{\sum_{i=1}^{N}(\widetilde{\rho}_{1,i}-1)^2}$ measures the contribution of the $1\%$ of tetrahedral elements with the largest absolute deviations $\lvert\widetilde{\rho}_{1,i}-1\rvert$ (indexed by $I_\rho$) to the total squared deviation of the normalized final density. It can be observed that $C_{1\%}^{\rho}$ is substantially lower for the free-boundary model across all three examples, indicating that the total squared density deviation is less dominated by the largest 1\% of elements. However, the IQR does not decrease uniformly across the examples. It is smaller for the Vase model, but larger for the Igea and David models. For the David model, the largest $1\%$ of density deviations still account for $60.21\%$ of the total squared deviation, indicating that the remaining density variation is substantially influenced by a small fraction of elements with large density deviations.
}

\begin{table}[t!]
\centering
\caption{Comparison of the three proposed models with 3DDEQCMap~\cite{lyu2025volumetric} for the Vase, Igea, and David models. For each model, we record {the number of iterations, the computational time,} the variance, {median, interquartile range (IQR) of the normalized final density $\widetilde{\rho}_{1} = \frac{\rho_1}{\text{Mean}(\rho_1)}$, $C_{1\%}^{\rho}$,} the mean and standard deviation of $K$, and the number of overlaps. The chosen/optimized radii for each model are also presented.}
\label{tab:three_model_comparison}

\resizebox{\linewidth}{!}{$
\begin{tabular}{llccccccccc}
\toprule
Example
& Model
& {\# Iterations}
& {Time (s)}
& $\operatorname{Var}(\widetilde{\rho}_{1})$
& {$\operatorname{Median}(\widetilde{\rho}_{1})$}
& {$\operatorname{IQR}(\widetilde{\rho}_{1})$}
& {$C_{1\%}^{\rho}$}
& $\operatorname{Mean}(K)$
& $\operatorname{SD}(K)$
& \# Overlaps\\
\midrule

\multirow{4}{*}{\shortstack[c]{Vase\\(14402)}}
& {3DDEQCMap}~\cite{lyu2025volumetric} $(1,1,1)$ & {18} & {4.3831}
& 0.8855 & {0.9121} & {0.1039} & {0.9827} & 2.4022 & 2.6903 & 0  \\
& Prescribed ellipsoid $(1,1,1.2)$ & {18} & {4.5349}
& 0.7321 & {0.9466} & {0.1004} & {0.9805} & 2.3366 & 2.9153 & 0 \\
& Adaptive ellipsoid $(1,0.9958,1.2050)$& {23} & {7.0081}
& 0.6108 & {0.9471} & {0.1005} & {0.9766} & 2.3244 & 2.2324 & 0 \\
& Free-boundary & {39} & {17.0929}
& 0.0705 & {0.9502} & {0.0804} & {0.5005} & 1.9849 & 0.7821 & 0 \\
\midrule

\multirow{4}{*}{\shortstack[c]{Igea\\(19539)}}
& {3DDEQCMap}~\cite{lyu2025volumetric} $(1,1,1)$ & {18} & {3.3337}
& 0.0480 & {0.9929} & {0.0517} & {0.9267} & 1.9176 & 0.8725 & 0 \\
& Prescribed ellipsoid $(1,1.2,1)$ & {20} & {3.2106}
& 0.0511 & {0.9932} & {0.0511} & {0.9381} & 1.8795 & 0.8053 & 0 \\
& Adaptive ellipsoid $(0.8620,1.1610,1.1991)$ & {26} & {8.1932}
& 0.0458 & {0.9933} & {0.0513} & {0.9285} & 1.6876 & 0.7971 & 0  \\
& Free-boundary & {29} & {18.7828}
& 0.0286 & {0.9824} & {0.0692} & {0.3708} & 1.7504 & 0.5063 & 0 \\
\midrule

\multirow{4}{*}{\shortstack[c]{David\\(23098)}}
& {3DDEQCMap}~\cite{lyu2025volumetric} $(1,1,1)$ & {17} & {11.0871}
& 0.7727 & {0.9525} & {0.0973} & {0.9689} & 2.3921 & 2.3228 & 0 \\
& Prescribed ellipsoid $(1,1,1.4)$ & {22} & {15.9386}
& 0.8781 & {0.9483} & {0.0967} & {0.9834} & 2.1979 & 2.2543 & 0 \\
& Adaptive ellipsoid $(0.9537,1.0129,1.4493)$ & {29} & {24.7900}
& 0.6537 & {0.9519} & {0.0935} & {0.9771} & 2.1827 & 2.0505 & 0 \\
& Free-boundary & {39} & {31.9254}
& 0.5802 & {0.9577} & {0.1056} & {0.6021} & 2.1087 & 1.5561 & 0 \\

\bottomrule
\end{tabular}
$}
\end{table}

\begin{figure}[t!]
    \centering
    \includegraphics[width=1\textwidth]{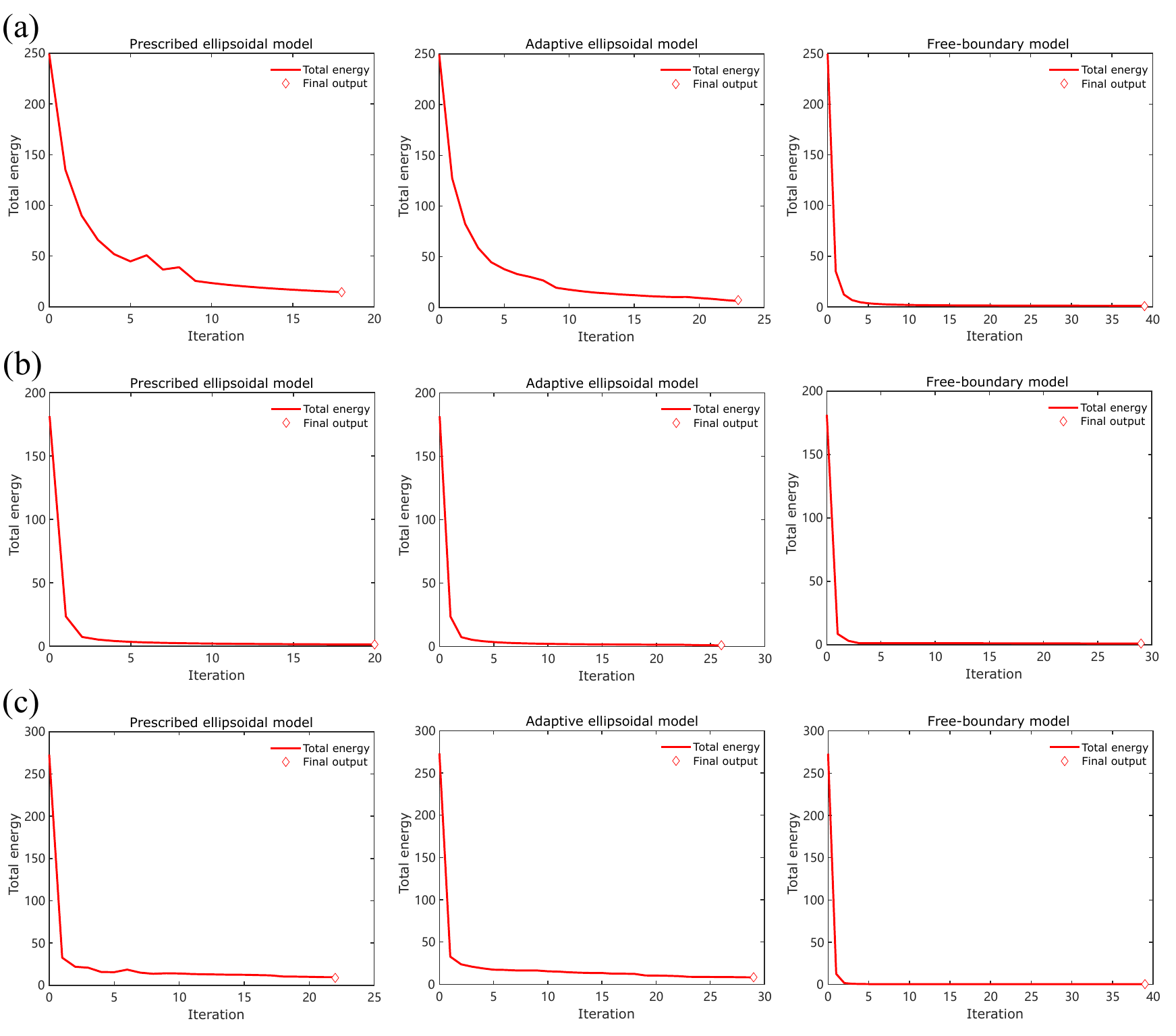}
    \caption{{\textbf{Energy charts of the three proposed volumetric parameterization models.} (a) Vase. (b) Igea. (c) David. From left to right: prescribed ellipsoidal model, adaptive ellipsoidal model, and free-boundary model. The red diamonds indicate the final outputs.}}
    \label{fig:example_energy}
\end{figure}

We next show the total energy after each complete iteration for the Vase, Igea, and David models under the three target-domain settings. Fig.~\ref{fig:example_energy} shows the total energy versus iteration. For all three examples, the total energy decreases substantially from its initial value. Most of the reduction occurs during the initial iterations, followed by a more gradual decrease. In particular, the free-boundary model exhibits a rapid initial energy reduction, after which the energy remains at a relatively low level. The prescribed and adaptive ellipsoidal models also show an overall decreasing trend, although small local increases occur in some iterations. These results illustrate the overall decrease in total energy during the coupled optimization process, without implying strict monotonicity at every iteration.

\subsection{Comparison with baseline methods}\label{sec:comparison_methods}
In this section, we compare our proposed methods with different baseline methods. In particular, we first compare our method with affine-rescaled 3DDEQCMap~\cite{lyu2025volumetric}. We then also compare our three proposed models with the volumetric harmonic map~\cite{wang2003volumetric} and optimal mass transport (OMT)~\cite{su2017volume} methods to evaluate their performance.

\subsubsection{Affine-rescaling comparison}
We apply anisotropic affine transformations to the unit-ball mappings produced by 3DDEQCMap~\cite{lyu2025volumetric} to obtain the corresponding ellipsoidal mappings, and compare them with our methods using the same target ellipsoids. Table~\ref{tab:rescaling_comparison} reports the variance of the normalized density as well as the mean and standard deviation of $K$. Note that affine rescaling does not change the variance of the normalized density, since it scales the volume of every tetrahedral element by the same factor while preserving its prescribed mass. However, anisotropic scaling generally changes the local shape distortion. Since the target radii are prescribed a priori, the resulting balance between density and shape distortions depends on how well the chosen ellipsoid matches the input geometry. In contrast, the adaptive ellipsoidal model jointly optimizes the mapping and target radii, consistently achieving lower density variance, mean $K$, and standard deviation of $K$ than the corresponding affine-rescaled baseline across all three test cases.

\begin{table}[t]
    
    \centering
    \caption{{\textbf{ Comparison of affine-rescaled 3DDEQCMap~\cite{lyu2025volumetric} and our ellipsoidal parameterization methods.} For each example, the rescaled mapping and the corresponding proposed method share the same target ellipsoid. We report the variance of the normalized density $\tilde{\rho}$, and the mean and standard deviation of the 3D quasi-conformal dilation $K$.}}
    \label{tab:rescaling_comparison}
    \small
    \begin{tabular}{lllccc}
        \toprule
        Example & Target radii $(a,b,c)$ & Method
        & $\operatorname{Var}(\tilde{\rho})$
        & Mean($K$) & SD($K$) \\
        \midrule
        Vase
        & $(1,1,1.2)$
        & Prescribed ellipsoid
        & 0.7321 & 2.3366 & 2.9153 \\
        & & Rescaled 3DDEQCMap
        & 0.8855 & 2.3775 & 2.8830 \\

        \addlinespace
        & $(1,0.9958,1.2050)$
        & Adaptive ellipsoid
        & 0.6108 & 2.3244 & 2.2324 \\
        & & Rescaled 3DDEQCMap
        & 0.8855 & 2.3790 & 2.8774 \\
        \midrule
        Igea
        & $(1,1.2,1)$
        & Prescribed ellipsoid
        & 0.0511 & 1.8795 & 0.8053 \\
        & & Rescaled 3DDEQCMap
        & 0.0480 & 1.9778 & 0.8720 \\

        \addlinespace
        & $(0.8620,1.1610,1.1991)$
        & Adaptive ellipsoid
        & 0.0458 & 1.6876 & 0.7971 \\
        & & Rescaled 3DDEQCMap
        & 0.0480 & 1.7132 & 0.8416 \\

        \midrule
        David
        & $(1,1,1.4)$
        & Prescribed ellipsoid
        & 0.8781 & 2.1979 & 2.2543 \\
        & & Rescaled 3DDEQCMap
        & 0.7727 & 2.4525 & 2.5345 \\

        \addlinespace
        & $(0.9537,1.0129,1.4493)$
        & Adaptive ellipsoid
        & 0.6537 & 2.1827 & 2.0505 \\
        & & Rescaled 3DDEQCMap
        & 0.7727 & 2.4714 & 2.4162 \\

        \bottomrule
    \end{tabular}
\end{table}

\subsubsection{Comparison with volumetric harmonic map and optimal mass transport}
\begin{figure}[t!]
    \centering
    \includegraphics[width=1\textwidth]{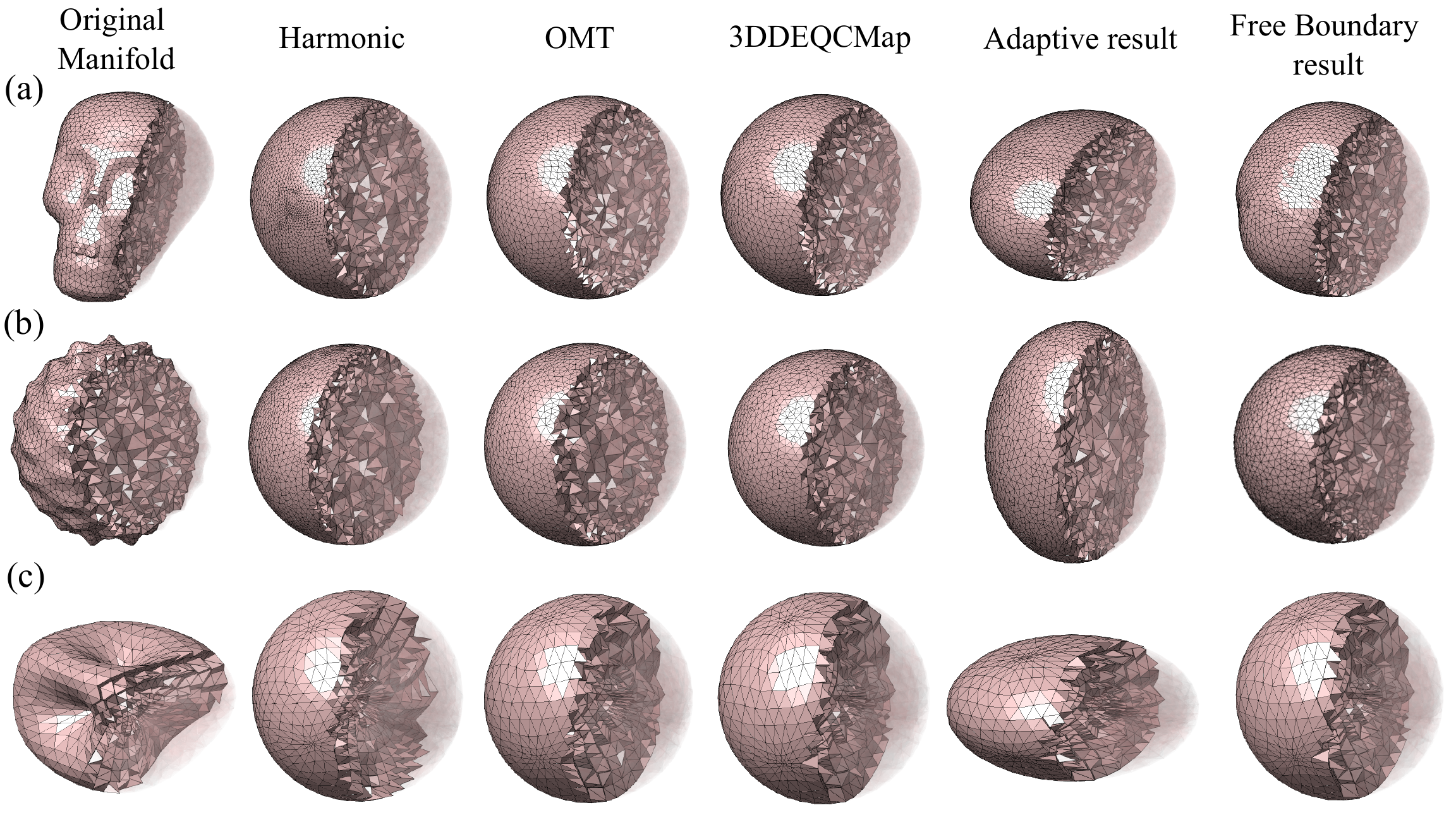}
    \caption{{\textbf{Comparison of volumetric parameterization results obtained by volumetric harmonic map~\cite{wang2003volumetric}, optimal mass transport (OMT)~\cite{su2017volume}, and our proposed methods.}  (a) Skull. (b) Multi-lobed ball. (c) Indented Sphere. From left to right: the original 3-manifold, the volumetric harmonic map result, the OMT result, the fixed-ball result, the adaptive ellipsoidal result, and the free-boundary result.}}
    \label{fig:comparison_methods}
\end{figure}

We also compare our volumetric parameterization framework with the volumetric harmonic map~\cite{wang2003volumetric} and the optimal mass transport (OMT) method~\cite{su2017volume}. We consider three examples: the Skull, Multi-lobed ball, and Indented Sphere. For each tetrahedron $T$, we prescribe the mass $m_T=\operatorname{Vol}(T)$ to encourage preservation of relative tetrahedral volumes. Fig.~\ref{fig:comparison_methods} shows the resulting volumetric parameterizations, including the original manifolds, the volumetric harmonic map~\cite{wang2003volumetric} and OMT~\cite{su2017volume} results, and those obtained by our three proposed models. Note that the volumetric harmonic map~\cite{wang2003volumetric}, OMT~\cite{su2017volume}, and 3DDEQCMap~\cite{lyu2025volumetric} use the unit ball as the target domain. Our adaptive ellipsoidal and free-boundary models also start from the unit-ball parameterization but allow the target domain to evolve during optimization. The adaptive ellipsoidal model adjusts the ellipsoid radii under the fixed-volume constraint, while the free-boundary model further relaxes the ellipsoidal constraint to allow non-ellipsoidal boundary deformations. These models provide increasing flexibility in target-domain geometry.

\begin{table}[t]
    
    \centering
    \small
    \setlength{\tabcolsep}{4pt}
    \caption{{\textbf{Comparison of volumetric harmonic map~\cite{wang2003volumetric} and optimal mass transport (OMT)~\cite{su2017volume} methods with 3DDEQCMap~\cite{lyu2025volumetric} and the proposed adaptive volumetric parameterization models.} For each example, we report the number of tetrahedral elements, the variance of the normalized density $\widetilde{\rho}$, the mean and standard deviation of the 3D quasi-conformal dilation $K$, and the number of overlaps.}}
    \label{tab:method_comparison}

    \resizebox{\linewidth}{!}{%
    \begin{tabular}{lclcccc}
        \toprule
        Example & \# Elements & Method
        & $\operatorname{Var}(\widetilde{\rho})$
        & Mean($K$) & SD($K$) & \# Overlaps \\
        \midrule

        \multirow{5}{*}{Skull}
        & \multirow{5}{*}{41938}
        & Volumetric Harmonic Map~\cite{wang2003volumetric} & 23.3132 & 1.5147 & 2.5960 & 1 \\
        & & OMT~\cite{su2017volume}
        & 0.3521 & 2.8812 & 2.2528 & 9 \\
        & & 3DDEQCMap~\cite{lyu2025volumetric}
        & 0.6843 & 1.8211 & 1.3704 & 0 \\
        & & Adaptive Ellipsoid
        & 0.6172 & 1.7965 & 1.3069 & 0 \\
        & & Free Boundary
        & 0.2706 & 1.6949 & 0.8247 & 0 \\

        \addlinespace

        \multirow{5}{*}{Multi-lobed ball}
        & \multirow{5}{*}{38917}
        & Volumetric Harmonic Map~\cite{wang2003volumetric} & 76.6965 & 2.3355 & 4.1159 & 1 \\
        & & OMT~\cite{su2017volume}
        & 171.9262 & 2.5463 & 16.4257 & 3 \\
        & & 3DDEQCMap~\cite{lyu2025volumetric}
        & 0.1248 & 1.3325 & 0.6724 & 0 \\
        & & Adaptive Ellipsoid
        & 0.1247 & 1.5597 & 0.6315 & 0 \\
        & & Free Boundary
        & 0.0505 & 1.2733 & 0.3787 & 0 \\

        \addlinespace

        \multirow{5}{*}{Indented Sphere}
        & \multirow{5}{*}{28160}
        & Volumetric Harmonic Map~\cite{wang2003volumetric} & 1.5073 & 2.1987 & 1.2285 & 0 \\
        & & OMT~\cite{su2017volume}
        & 30.7363 & 4.2869 & 25.6475 & 444 \\
        & & 3DDEQCMap~\cite{lyu2025volumetric}
        & 0.1761 & 3.2306 & 1.7598 & 0 \\
        & & Adaptive Ellipsoid
        & 0.1892 & 3.6353 & 2.0541 & 0 \\
        & & Free Boundary
        & 0.0570 & 2.4068 & 0.9575 & 0 \\

        \bottomrule
    \end{tabular}%
    }
\end{table}

Table~\ref{tab:method_comparison} reports the variance of the normalized density, the mean of $K$, its standard deviation, and the number of overlaps for the five methods. It can be observed that the volumetric harmonic map~\cite{wang2003volumetric} and OMT methods~\cite{su2017volume} exhibit substantially larger density variances for the Multi-lobed Ball and Indented Sphere models. In particular, OMT~\cite{su2017volume} produces a large number of overlaps for the Indented Sphere. Among 3DDEQCMap~\cite{lyu2025volumetric} and the two adaptive models, the free-boundary model consistently achieves the lowest density variance, mean $K$, and standard deviation of $K$ across all three examples. Moreover, both the adaptive ellipsoidal and free-boundary models produce overlap-free mappings in all three cases. These results demonstrate the advantages of target-domain adaptation in achieving a better balance between density equalization and local shape distortion.

{
\subsection{Sensitivity Analysis}
In this section, we further investigate the sensitivity of the proposed methods to initialization and the size of the auxiliary sea.

\subsubsection{Initialization sensitivity}

\begin{table}[t]
    
    \centering
    \small
    \caption{{\textbf{Sensitivity of the adaptive ellipsoidal model to initialization.} For the Igea model, we consider seven different initial ellipsoidal radii and volume factors $P_0=a_0b_0c_0$.
    We report the final radii, the variance of the normalized final density $\widetilde{\rho}_1$, the mean and standard deviation of the 3D quasi-conformal dilation $K$, and the number of overlaps.}}
    \label{tab:adaptive_initialization}
    \begin{tabular}{cccccc}
        \hline
        Initial radii $(a_0,b_0,c_0)$ &
        Final radii $(a_1,b_1,c_1)$ &
        Var($\widetilde{\rho}_1$) &
        Mean($K$) &
        SD($K$) &
        \# Overlaps \\
        \hline
        $(1,1,1)$   & $	(0.8891, 1.1248, 1.000)$ & 0.0637 & 1.7976 & 0.8411 & 0 \\
        $(1,1,0.8)$ & $	(0.8220, 0.9608, 1.0129)$ & 0.0534 & 1.7313 & 0.7952 & 0 \\
        $(1,1,1.2)$ & $	(0.8521, 1.1735, 1.2000)$ & 0.0762 & 1.6769 & 0.8153 & 0 \\
        $(1,1,1.4)$ & $	(0.8855, 1.2431, 1.2718)$ & 0.0968 & 1.6797 & 0.8684 & 0 \\
        $(1,0.8,1)$ & $	(0.8220, 0.9732, 1.0000)$ & 0.0773 & 1.7295 & 0.8386 & 0 \\
        $(1,1.2,1)$ & $(0.8620, 1.1610, 1.1991)$  & 0.0458 & 1.6876 & 0.7971 & 0 \\
        $(1,1.4,1)$ & $	(0.8855, 1.2487, 1.2662)$ & 0.0526 & 1.6764 & 0.7637 & 0 \\        
        \hline
    \end{tabular}
\end{table}
We first investigate the sensitivity of the adaptive ellipsoidal and free-boundary models to initialization using the Igea model. We consider seven different initial ellipsoids with varying radii and volumes.

\begin{table}[t]
    
    \centering
    \caption{{\textbf{Sensitivity of the free-boundary model to initialization.} For the Igea model, we consider seven different initial ellipsoidal radii. We report the variance of the normalized final density $\widetilde{\rho}_1$, the mean and standard deviation of the 3D quasi-conformal dilation $K$, and the number of overlaps.}}
    \label{tab:free_initialization}
    \begin{tabular}{ccccc}
        \hline
        Initial radii $(a_0,b_0,c_0)$ &
        Var($\widetilde{\rho}_1$) &
        Mean($K$) &
        SD($K$) &
        \# Overlaps \\
        \hline
        $(1,1,1)$   & 0.0237 & 1.8222 & 0.5353 & 0 \\
        $(1,1,0.8)$ & 0.0243 & 2.0609 & 0.5744 & 0 \\
        $(1,1,1.2)$ & 0.0227 & 1.7605 & 0.5197 & 0 \\
        $(1,1,1.4)$ & 0.0216 & 1.8309 & 0.5143 & 0 \\
        $(1,0.8,1)$ & 0.0224 & 2.0930 & 0.6251 & 0 \\
        $(1,1.2,1)$ & 0.0286 & 1.7504 & 0.5063 & 0 \\
        $(1,1.4,1)$ & 0.0239 & 1.8155 & 0.4547 & 0 \\
        \hline
    \end{tabular}
\end{table}

Table~\ref{tab:adaptive_initialization} shows that the adaptive ellipsoidal model yields similar mean and standard deviation of $K$ under different initializations. The final density variance exhibits more pronounced changes. In particular, it increases with the initial radius along the $z$-axis when the other radii are fixed. In contrast, Table~\ref{tab:free_initialization} shows that the free-boundary model maintains relatively stable density variance across different initializations. Its local shape distortion is more sensitive to the initial radii, especially when the radius along the $y$- or $z$-axis is reduced. The additional boundary flexibility may help maintain density equalization, although the initial geometry still affects local shape distortion. Compared with the adaptive ellipsoidal model, the free-boundary model achieves lower density variance and standard deviation of $K$, but higher mean $K$, for all seven initializations. Both models produce folding-free mappings with zero overlaps.

\subsubsection{Sea size sensitivity}
We next test how auxiliary sea size affects the free-boundary model using the Vase model. In all tests, the free-boundary model is initialized with an ellipsoidal parameterization, with initial radii $(a_0,b_0,c_0)=(1,1,1.2)$. Tables~\ref{tab:sea_layer_sensitivity} and~\ref{tab:sea_thickness_sensitivity} show results obtained by varying the number of sea layers $L$ and the layer-thickness parameter $\kappa$, respectively, while fixing the other parameter.

\begin{table}[t]
    
    \centering
    \small
    \caption{{\textbf{Sensitivity of the free-boundary model to the number of sea layers.} For the Vase model, we consider different numbers of auxiliary sea layers $L$, with the layer thickness parameter fixed at $\kappa=2$. We report the variance of the normalized final density $\widetilde{\rho}_1$, the mean and standard deviation of the 3D quasi-conformal dilation $K$, and the number of overlaps.}}
    \label{tab:sea_layer_sensitivity}
    \begin{tabular}{ccccc}
        \hline
        Sea layers $L$ &
        Var($\widetilde{\rho}_1$) &
        Mean($K$) &
        SD($K$) &
        \# Overlaps \\
        \hline
        1 & 0.1476 & 2.1040 & 0.9801 & 0 \\
        2 & 0.0752 & 2.0115 & 0.8079 & 0 \\
        3 & 0.0705 & 1.9849 & 0.7821 & 0 \\
        4 & 0.0712 & 1.9741 & 0.7766 & 0 \\
        5 & 0.0709 & 1.9716 & 0.7759 & 0 \\
        6 & 0.0692 & 1.9668 & 0.7761 & 0 \\
        \hline
    \end{tabular}
\end{table}

\begin{table}[t]
    
    \centering
    \small
    \caption{{\textbf{Sensitivity of the free-boundary model to the layer thickness parameter.}
    For the Vase model, we consider different values of $\kappa$ with the number of auxiliary sea layers fixed at $L=3$.
    We report the variance of the normalized final density $\widetilde{\rho}_1$, the mean and standard deviation of the 3D quasi-conformal dilation $K$, and the number of overlaps.}}
    \label{tab:sea_thickness_sensitivity}
    \begin{tabular}{ccccc}
        \hline
        Parameter $\kappa$ &
        Var($\widetilde{\rho}_1$) &
        Mean($K$) &
        SD($K$) &
        \# Overlaps \\
        \hline
        1.0 & 0.1185 & 1.9814 & 0.9797 & 0 \\
        1.5 & 0.0855 & 1.9745 & 0.8114 & 0 \\
        2.0 & 0.0705 & 1.9849 & 0.7821 & 0 \\
        2.5 & 0.0765 & 1.9991 & 0.7749 & 0 \\
        3.0 & 0.0689 & 2.0148 & 0.7801 & 0 \\
        3.5 & 0.0758 & 1.9977 & 0.8297 & 0 \\
        \hline
    \end{tabular}
\end{table}
From Table~\ref{tab:sea_layer_sensitivity}, we can observe that increasing the number of sea layers initially reduces both density and local shape distortions. When $L$ exceeds 3, the density variance shows only minor changes, and the improvements in shape distortion are also very limited. This indicates that a sufficiently large auxiliary sea area provides additional flexibility for density diffusion and boundary deformation. However, further increasing its size yields diminishing benefits. For the layer thickness parameter, Table~\ref{tab:sea_thickness_sensitivity} shows that the density variance decreases initially as $\kappa$ increases, but does not improve consistently for larger values. Meanwhile, the mean of $K$ remains nearly unchanged, while its standard deviation decreases initially and then increases slightly for larger values of $\kappa$. These results suggest that the default setting $L=3$ and $\kappa=2$ provides a reasonable choice for balancing density and local shape distortions without unnecessarily enlarging the auxiliary sea. All resulting mappings are folding-free, with zero overlaps.

}

\section{Applications}\label{sec:applications}
In this section, we showcase the applicability of our proposed volumetric parameterization framework for 3-manifold remeshing, registration, and morphing.

\subsection{3-manifold Remeshing}
Our proposed framework can be easily applied to 3-manifold remeshing. Specifically, given a 3D simply connected manifold $\mathcal{M}$ in $\mathbb R^3$, we first compute an ellipsoidal volumetric parameterization $f:\mathcal{M}\to \mathbb{E}_{a,b,c}$. Then, we can easily generate a uniform tetrahedral mesh in the ellipsoidal domain using existing tools such as DistMesh~\cite{persson2004simple}. Finally, utilizing the inverse mapping $f^{-1}$, we can map the uniform mesh onto $\mathcal{M}$, yielding a remeshed 3-manifold.

\begin{figure}[t!]
    \centering
    \includegraphics[width=0.95\textwidth]{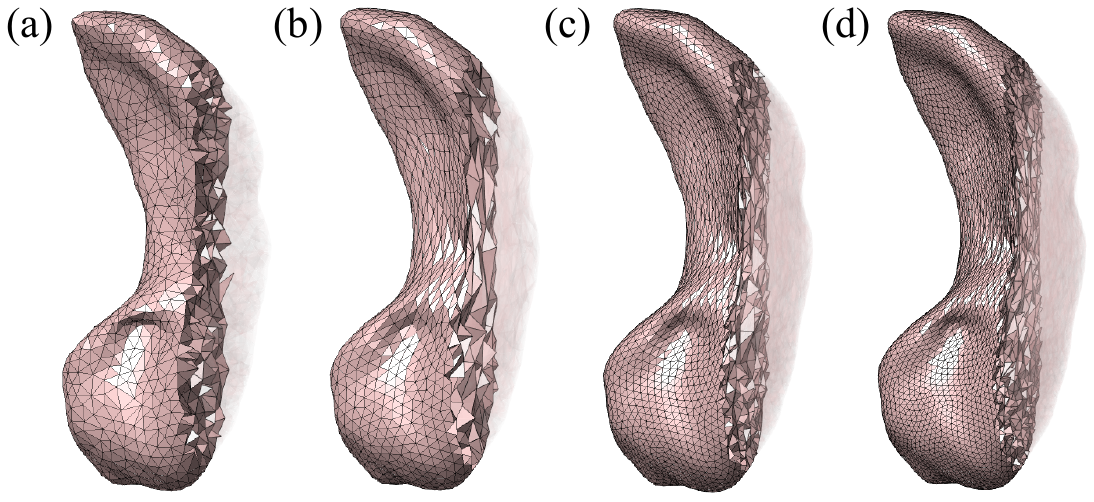}
    \caption{\textbf{Multiresolution volumetric remeshing of the Hippocampus model.} (a) The original Hippocampus model.  (b)--(d) The remeshed results with approximately 4K, 8K, and 15K vertices, respectively. }
    \label{fig:hipp_remesh}
\end{figure}

In Fig.~\ref{fig:hipp_remesh}, we present several examples of remeshing of the brain hippocampus model with different numbers of vertices. Compared with the original model, the remeshed manifolds exhibit high mesh quality, with uniform and regular tetrahedral elements. More specifically, as the {resolution} increases, the features of the remeshed hippocampus become clearer, especially in the high-curvature domain. 

\begin{figure}[t!]
    \centering
    \includegraphics[width=1\textwidth]{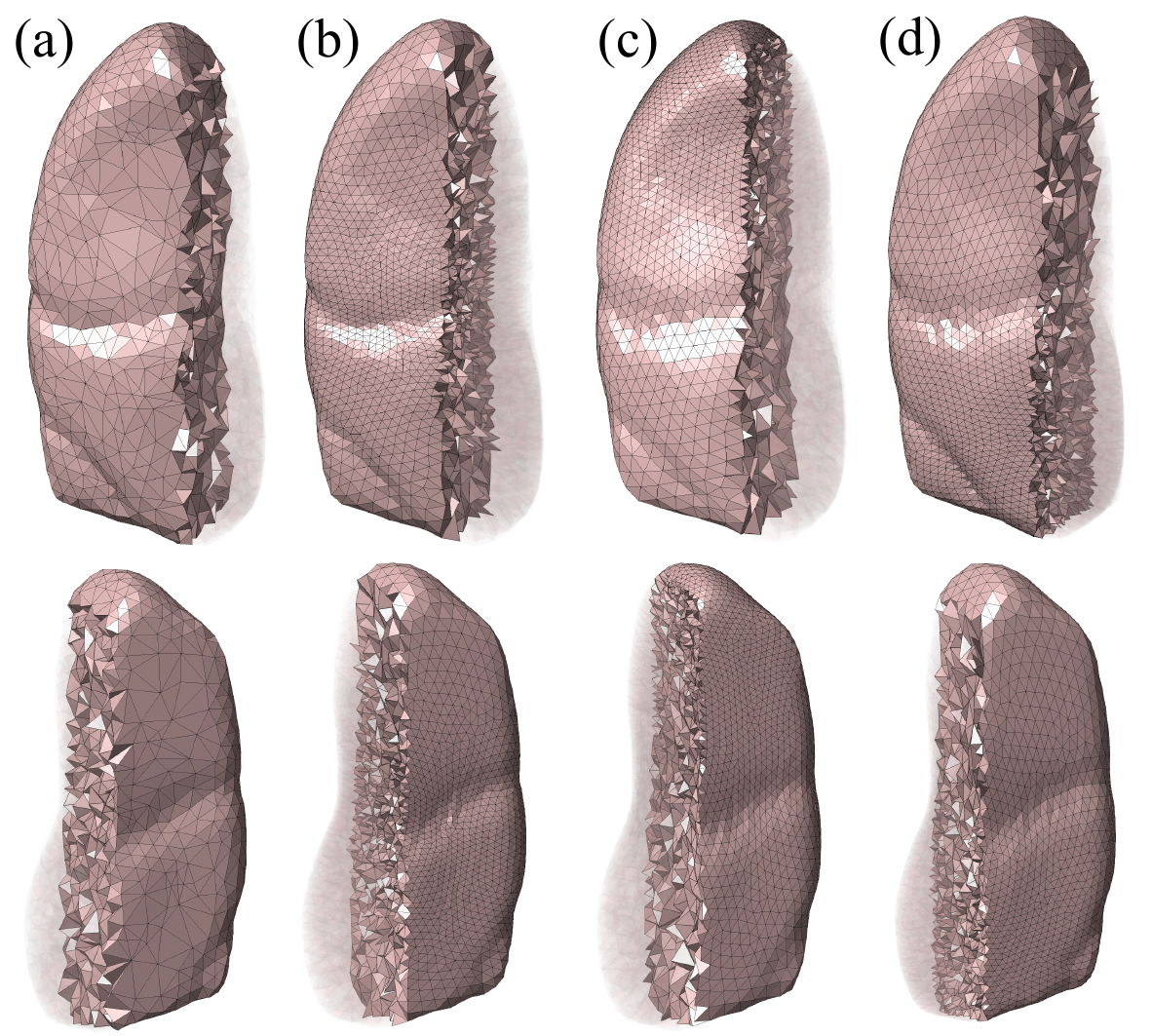}
    \caption{\textbf{Localized adaptive volumetric remeshing of the Lung model.} (a) The original Lung model. (b)--(d) show the adaptive mesh reconstruction results after local mesh refinement in the central, top, and bottom regions of the lung, respectively, simulating different possible locations of lung lesions.}
    \label{fig:Lung_adaptive}
\end{figure}

Note that the effect of the proposed methods is closely related to the prescribed mass distribution. Specifically, the high-density domain would be enlarged while the low-density domain would be shrunk. Therefore, we can obtain adaptive remeshing results by setting different mass distributions. Fig.~\ref{fig:Lung_adaptive} shows several adaptive remeshing results of the Lung model to demonstrate this localized adaptivity. Here we consider a Lung model and simulate possible lesions located in different anatomical regions. Specifically, we set prescribed mass distributions concentrated in the central, top, and bottom regions. The original Lung model is shown in Fig.~\ref{fig:Lung_adaptive}(a). Fig.~\ref{fig:Lung_adaptive}(b)--(d) present the remeshing results with different user-defined mass distributions. It can be observed that the mesh density in these special regions is higher than in the other parts.

{
To quantitatively evaluate tetrahedral remeshing quality, we consider both element-size and shape variations. Let $\mathcal{T}=\{K_i\}_{i=1}^{N_T}$ be a tetrahedral mesh, where $V_i>0$ and $\ell_{i,1},\ldots,\ell_{i,6}$ denote the volume and six edge lengths of $K_i$, respectively.

The element-size variation is defined as
\begin{equation}
    \delta_{\mathrm{size}}^{\mathrm{tet}} = \log\left(\frac{V_{\max}}{V_{\min}}\right),
    \label{eq:tet_size_variation}
\end{equation}
where $V_{\max}=\max_i V_i$ and $V_{\min}=\min_i V_i$. Clearly, $\delta_{\mathrm{size}}^{\mathrm{tet}}\geq0$, with equality if and only if all tetrahedral elements have the same volume.

For shape variation, we define an edge-length deviation measure for each element and its average over the mesh:
\begin{equation}
    R_i^{\mathrm{tet}} = \sum_{j=1}^{6}\left|\frac{\ell_{i,j}}{\sum_{k=1}^{6}\ell_{i,k}}-\frac16\right|, \qquad
    \delta_{\mathrm{shape}}^{\mathrm{tet}} = \frac1{N_T}\sum_{i=1}^{N_T}R_i^{\mathrm{tet}}.
    \label{eq:tet_shape_variation}
\end{equation}
Here, $R_i^{\mathrm{tet}}=0$ if and only if $K_i$ is regular. Consequently, $\delta_{\mathrm{shape}}^{\mathrm{tet}}\geq0$, with equality if and only if all elements are regular. Smaller values indicate more uniform element volumes and shapes, respectively. For localized adaptive remeshing, however, nonuniform element sizes are intentional, so $\delta_{\mathrm{size}}^{\mathrm{tet}}$ describes the range of element volumes rather than a quantity to be minimized.

Since comparable edge lengths do not exclude sliver tetrahedra with nearly coplanar vertices, the edge-based shape measure alone cannot fully characterize volumetric degeneracy. We therefore also consider the standard mean-ratio quality~\cite{LiuJoe1994}:
\begin{equation}
    q_i = \frac{12(3V_i)^{2/3}}{\sum_{j=1}^{6}\ell_{i,j}^{\,2}}.
    \label{eq:tet_mean_ratio}
\end{equation}
For a nondegenerate tetrahedron, $0<q_i\leq1$, with $q_i=1$ if and only if $K_i$ is regular. Moreover, $q_i\to0$ as $V_i/h_i^3\to0$, where $h_i=\max_j\ell_{i,j}$ is the element diameter, allowing it to detect degeneracy, including sliver configurations. All three measures are dimensionless and invariant under rigid motions and uniform scaling. We report the mean and standard deviation of $q_i$ to characterize the average element quality and its variability, respectively, with larger mean values indicating better average quality.

\begin{table}[t]
    
    \centering
    \caption{{\textbf{Quality assessment of tetrahedral remeshing results.} We report the number of tetrahedral elements $N_T$, the size and shape variations $\delta_{\mathrm{size}}^{\mathrm{tet}}$, $\delta_{\mathrm{shape}}^{\mathrm{tet}}$, and the mean and standard deviation of the mean-ratio quality over all tetrahedral elements Mean($q_i$), SD($q_i$).}}
    \label{tab:tet_remeshing_quality}
    \begin{tabular}{llccccc}
        \toprule
        Example & Mesh setting & $N_T$
        & $\delta_{\mathrm{size}}^{\mathrm{tet}}$
        & $\delta_{\mathrm{shape}}^{\mathrm{tet}}$
        & Mean($q_i$) & SD($q_i$) \\
        \midrule
        Hippocampus
        & 4K & 10623
        & 7.2737 & 0.2607 & 0.5396 & 0.2034 \\
        & 8K & 40640
        & 7.7561 & 0.2600 & 0.5450 & 0.2031 \\
        & 15K & 73279
        & 8.2325 & 0.2616 & 0.5442 & 0.2026 \\
        \midrule
        Lung
        & Center & 40674
        & 5.2868 & 0.1979 & 0.7159 & 0.1427 \\
        & Top & 40674
        & 5.6745 & 0.1735 & 0.7474 & 0.1339 \\
        & Bottom & 40674
        & 6.2594 & 0.1811 & 0.7340 & 0.1377 \\
        \bottomrule
    \end{tabular}
\end{table}

Table~\ref{tab:tet_remeshing_quality} reports the values of $\delta_{\mathrm{size}}^{\mathrm{tet}}$ and $\delta_{\mathrm{shape}}^{\mathrm{tet}}$, together with the mean and standard deviation of $q_i$, for all remeshing results shown in Figs.~\ref{fig:hipp_remesh} and \ref{fig:Lung_adaptive}. For the Hippocampus model, the shape variation remains nearly constant across the three resolutions, and both the mean and standard deviation of $q_i$ are also stable. In contrast, the size variation increases with resolution, indicating a wider range of element volumes. These results suggest that the average element-shape quality and its variability remain comparable across different resolutions, although the uniformity of element sizes decreases. For the Lung model, the differences in size variation reflect the prescribed localized refinement patterns. The shape variation, mean values, and standard deviations of $q_i$ are comparable across the three results, indicating similar average element-shape quality and variability under different localized refinement patterns.

}

\subsection{3-manifold Registration and Morphing}
The proposed volumetric parameterization provides a common canonical domain for establishing volumetric correspondences between simply connected 3-manifolds. Let $\mathcal M_0$ and $\mathcal M_1$ be two 3-manifolds. Using the prescribed ellipsoidal algorithm, we can compute the volumetric parameterizations $f_0:\mathcal{M}_0 \to \mathbb E_{a,b,c}$ and $f_1:\mathcal{M}_1 \to \mathbb E_{a,b,c}$ with fixed radii $a,b,c$. Here, we set $a = b$ in the solid ellipsoidal domain $\mathbb E_{a,b,c}$.

\begin{figure}[t!]
    \centering
    \includegraphics[width=0.95\textwidth]{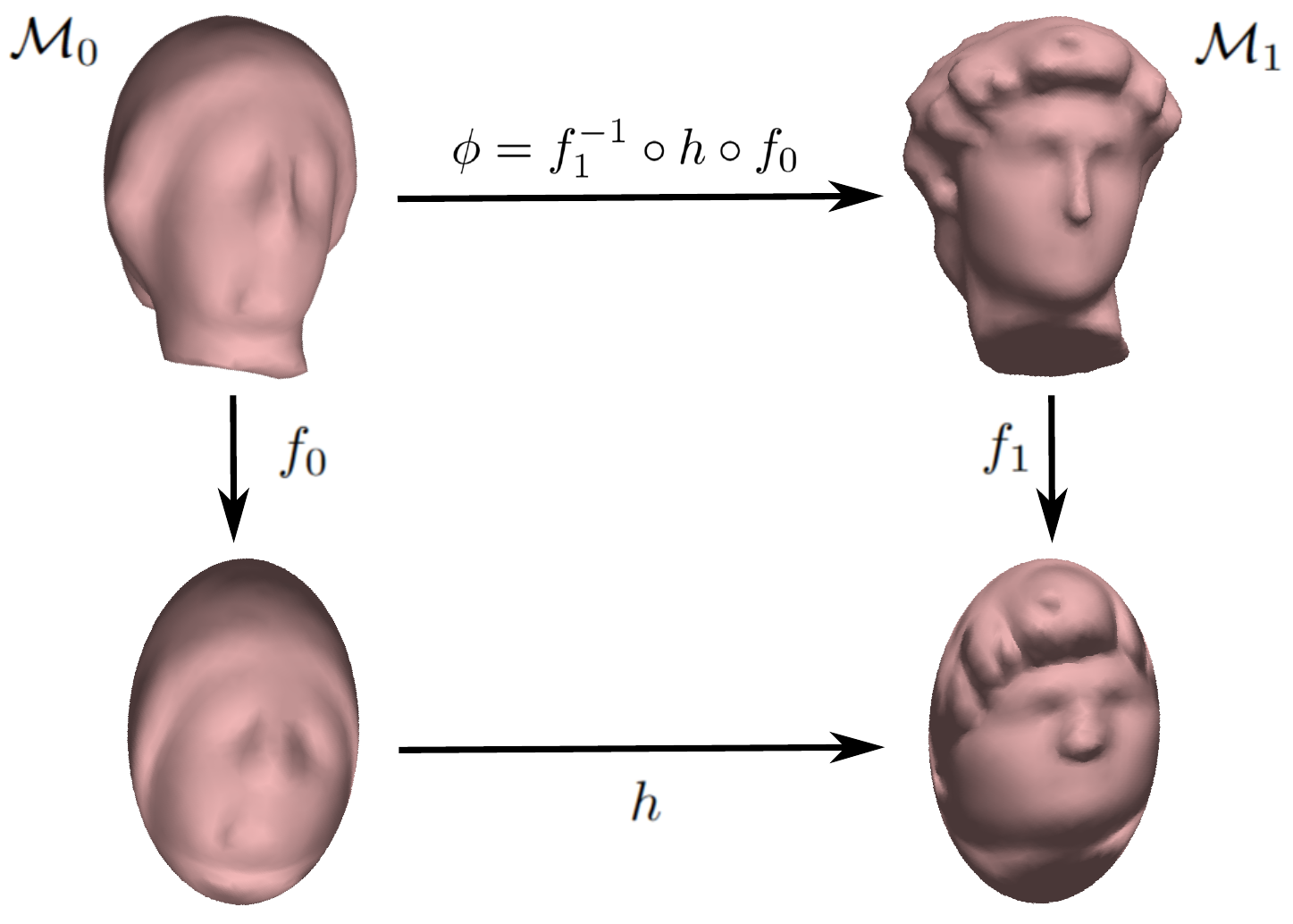}
    \caption{\textbf{Registration between simply connected 3-manifolds via our proposed volumetric parameterization methods.} The source and target volumetric meshes $\mathcal{M}_0$ and $\mathcal{M}_1$ are mapped onto the same prescribed solid ellipsoid by the parameterizations $f_0$ and $f_1$, respectively. After aligning their principal $z$-axes, the remaining rotational freedom is removed by an optimized rotation. The resulting volumetric registration map is given by $\phi=f_1^{-1}\circ h\circ f_0$. }
    \label{fig:Reg}
\end{figure}

Since the ellipsoid is rotationally symmetric about the principal $z$-axis, we can remove the rotational degree of freedom by solving
\begin{equation}
    \theta_i = \underset{\theta\in[0,2\pi)}{\operatorname{argmin}}\int_{\mathcal M_i}\|R_z(\theta)f_i(x)-\widehat{\iota}_i(x)\|_2^2,
    \qquad i=0,1,
\end{equation}
where $\widehat{\iota}_i$ denotes the centered and scale-normalized
embedding of $\mathcal M_i$ and $R_z$ is the rotation function along $z$-axis. Then, we define $h = R_z(-\theta_1)R_z(\theta_0)=R_z(\theta_0-\theta_1).$ Consequently, the ellipsoidal volumetric registration map from $\mathcal M_0$ to $\mathcal M_1$ is defined as $\phi = f_1^{-1}\circ h\circ f_0.$ The illustration of the registration is given in Fig.~\ref{fig:Reg}.

\begin{figure}[t!]
    \centering
    \includegraphics[width=1\textwidth]{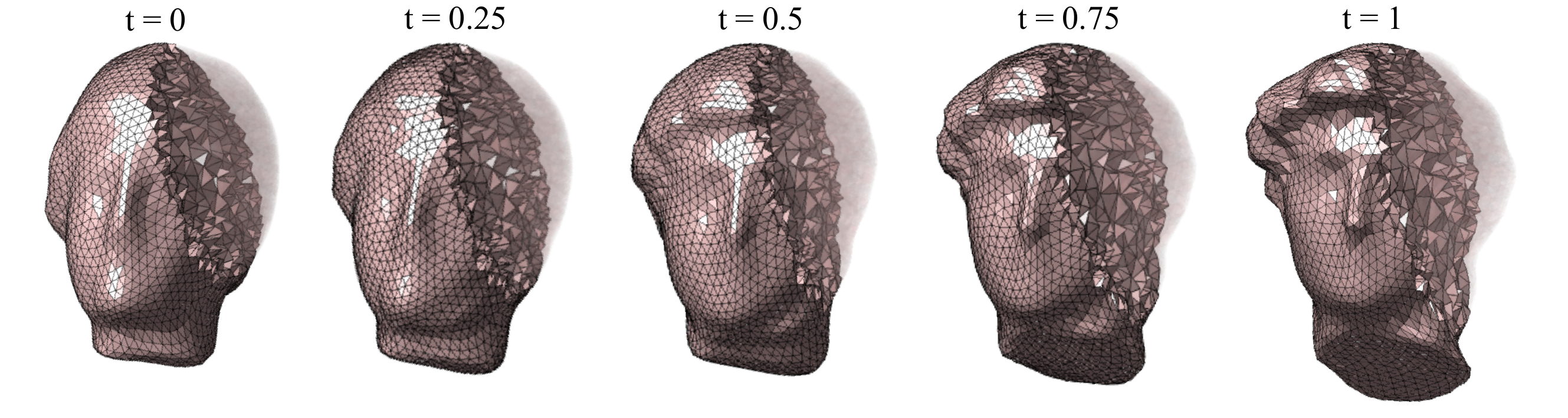}
    \caption{\textbf{3-manifold homotopy via our proposed volumetric parameterization methods.} Here, we construct a continuous deformation from the Igea model ($t=0$) to the David model ($t=1$). The intermediate volumetric mappings at $t=0,\,0.25,\,0.5,\,0.75,$ and $1$ illustrate the continuous transition between the two models. The sectional view shows the internal tetrahedral mesh.}
    \label{fig:deformation}
\end{figure}

Based on the obtained volumetric registration, we can further construct a continuous morphing sequence between $\mathcal{M}_0$ and $\mathcal{M}_1$. Specifically, the linear homotopy is defined as
\begin{equation} \label{eq:homotopy}
    \mathcal{L}(x,t) = (1-t)x + t \phi (x).
\end{equation}
It is clear that $\mathcal{L}(\mathcal{M}_0,0)=\mathcal{M}_0$ and
$\mathcal{L}(\mathcal{M}_0,1) = \mathcal{M}_1$. In Fig.~\ref{fig:deformation}, we present the morphing results between two volumetric meshes at $t=0$, $0.25$, $0.5$, $0.75$, and $1$. The intermediate maps exhibit a smooth and gradual transition in both the overall shape and the local geometric features. Altogether, the above examples demonstrate the applicability of our method for the registration and morphing of volumetric meshes.

To quantitatively evaluate the registration accuracy, we select five pairs of corresponding facial landmarks on the Igea and David models. These landmarks correspond to the left and right eye sockets, the nose tip, and the left and right mouth corners, respectively. They are used only for evaluation and are not involved in computing the registration map. Table~\ref{tab:registration_accuracy} reports the corresponding landmark errors. For each pair $(p_i,q_i)$, we define the pre-registration and post-registration errors on the common canonical ellipsoid, respectively, as
\begin{equation}
    \mathrm{LE}_{1,i}=\|f_0(p_i)-f_1(q_i)\|_2,
    \qquad
    \mathrm{LE}_{2,i}=\|h\circ f_0(p_i)-f_1(q_i)\|_2.
\end{equation}
The landmark errors decrease for all five pairs after registration, indicating improved correspondence on the common canonical ellipsoid.

\begin{table}[t]
    
    \centering
    \small
    \caption{{\textbf{Landmark correspondence errors on the canonical ellipsoid between the Igea and David models.} $\mathrm{LE}_1$ and $\mathrm{LE}_2$ denote the pre-registration and post-registration $L^2$ errors, respectively.}}
    \label{tab:registration_accuracy}
    \begin{tabular}{lccccc}
        \toprule
        & P1 & P2 & P3 & P4 & P5 \\
        \midrule
        Landmark
        & Left eye
        & Right eye
        & Nose
        & Left mouth
        & Right mouth \\
        \midrule
        $\mathrm{LE}_1$ & 1.9089 & 1.8477 & 1.9767 & 1.8248 & 1.8345 \\
        $\mathrm{LE}_2$ & 0.1183 & 0.2443 & 0.0878 & 0.2088 & 0.1139 \\
        \bottomrule
    \end{tabular}
\end{table}

\begin{table}[t!]
    
    \centering
    \small
    \caption{{\textbf{Performance of the Igea--David morphing.}
    We report the mean and standard deviation of the signed 3D quasi-conformal dilation $K$, and the number of overlaps at different sampled time points.}}
    \label{tab:morphing_injectivity}
    \begin{tabular}{lccccc}
        \toprule
        Time $t$
        &0 &  0.25 &  0.50 &  0.75 &  1\\
        \midrule
        Mean($K$)
        & 1&  1.2046 &  1.4704 &  1.9997 &  2.3730 \\
        SD($K$)
        &0 &  0.3949 &  0.5169 &  1.2927 &  2.8084 \\
        \# Overlaps
        & 0 & 0 & 0 & 0 & 0  \\
        \bottomrule
    \end{tabular}
\end{table}

It is noteworthy that the bijectivity of the endpoint registration does not guarantee injectivity at intermediate times. As a numerical check, we examine the linear morphing configurations at different sampled times between $t=0$ and $t=1$. At each tested time, we compute the signed 3D quasi-conformal dilation $K$ relative to the initial Igea mesh and report its mean and standard deviation, together with the number of overlaps between distinct tetrahedral elements. As shown in Table~\ref{tab:morphing_injectivity}, the number of detected overlaps is zero at each of the five sampled times. More generally, if folding is detected in an intermediate mesh, we can apply the geometric correction scheme described in Section~\ref{sec:geometry_correctopn_scheme} to the linear homotopy result in Eq.~\eqref{eq:homotopy}.

\section{Conclusion}\label{sec:conclusion}
In this work, we have proposed a novel framework for computing adaptive volumetric parameterization for simply connected 3-manifolds. Our framework simultaneously controls the local shape and mass distortions while optimizing the parametric domain. Depending on the target domain family, we develop three progressively more flexible models: a prescribed ellipsoidal model, a volume-normalized adaptive ellipsoidal model, and a sea-embedded free-boundary model. These models provide increasing flexibility in target geometry, from fixed ellipsoids to adaptively optimized ellipsoids, and even general non-ellipsoidal target domains, thereby allowing us to easily achieve different volumetric mappings while effectively controlling local geometric distortion. The experimental results have demonstrated the effectiveness of our proposed methods. In particular, our methods can be easily utilized for a large variety of structures ranging from digital models commonly used in computer graphics to medical shapes such as the skull, hippocampus, and lung. Altogether, our proposed methods provide a new approach for the representation and processing of simply connected 3-manifolds for practical applications in manifold remeshing, registration, and morphing.

Note that our proposed volumetric parameterization framework is currently limited to simply connected 3-manifolds. In the future, we plan to extend the methods for manifolds with more complex topologies. Besides, the current methods focus on volumetric parameterization with controllable local shape and mass distortions. A possible future direction is to incorporate landmark-matching constraints into our framework, which can be used for landmark-based 3-manifold registration. Finally, as our current methods are only applicable to tetrahedral meshes, another natural next step is to extend them to point clouds and hexahedral meshes.

\bibliographystyle{ieeetr}
\bibliography{VEMbib.bib}

\end{document}